\documentclass[aps, twocolumn, prb, longbibliography]{revtex4-2}
\usepackage{amsmath, amsfonts, amssymb, wasysym, mathrsfs, dsfont}
\usepackage{graphicx, color}
\usepackage{ulem} 
\usepackage{dcolumn} 
\usepackage[dvipsnames]{xcolor}
\usepackage{xspace}
\usepackage{nicefrac, bm}
\usepackage[colorlinks, breaklinks, linkcolor=OrangeRed, citecolor=RoyalBlue, urlcolor=NavyBlue]{hyperref}
\usepackage[all]{hypcap} 

\newcommand{\gsbra}{\left<\text{GS}\right|}
\newcommand{\gsket}{\left|\text{GS}\right>}
\newcommand{\tran}{\text{T}}
\newcommand{\ket}[1]{\left|#1\right>}
\newcommand{\bra}[1]{\left<#1\right|}
\newcommand{\fermi}{\ket{\text{F}}}
\newcommand{\sset}[2]{{\left\{#1_i\right\}_{#2}}}
\newcommand{\tk}{T_\text{K}}

\begin{document}

\title{The structure of quasiparticles in a local Fermi liquid}
\author{Izak Snyman}
\affiliation{Mandelstam Institute for Theoretical Physics, School of Physics, University of the Witwatersrand,
Johannesburg, South Africa}

\begin{abstract}
Conduction electrons interacting with a dynamic impurity can give rise to a local Fermi liquid. The latter
has the same  low energy spectrum as an ideal Fermi gas containing a static impurity.
The Fermi liquids's elementary excitations are however not bare electrons. 
In the vicinity of the impurity, they are dressed by virtual particle-hole pairs.
Here we study this dressing. Among other things, we construct a mode-resolved measure of dressing.  
To evaluate it in position representation, 
we have to circumvent the limitations of the Numerical Renormalization Group, which discretizes the conduction band logarithmically.
We therefore extend Natural Orbital methods, that successfully characterize the ground state,
to describe excitations. We demonstrate that the dressing profile shows nontrivial powerlaw decay at large distances. 
Our Natural Orbital methodology could lay the foundation for calculating the properties of local Fermi liquid quasiparticles in nontrivial geometries such
as disordered hosts or mesoscopic devices.
\end{abstract}
\maketitle

\section{Introduction}

Fermi liquids emerge near infrared fixed points 
in the Renormalization Group flow of some interacting many-fermion systems \cite{shankar}. 
There is something alchemical about them. Renormalization plays the role of the philosopher's stone,
so that it often remains a mystery how the independent quasiparticles (the probverbial gold) are constituted out of the raw ingredients 
-- bare electrons.
Physicists characterize the low energy behavior of Fermi liquids in terms of a handful of parameters that
are determined from experiment. 
A more daunting task is to determine how the
Fermi liquid parameters depend on the microscopic parameters of a given physical realization.

Local Fermi liquids are a class of systems in which this challenging problem has met with success \cite{hewsonbook}. 
They occur in dynamic quantum impurity models, where a small interacting quantum system is coupled to a bulk system of non-interacting fermions. 
They are interesting many-body systems in their own right \cite{Pustilnik}, and also appear as a key ingredient in dynamical mean field theory \cite{dmft}, an
important method in the study of bulk-interacting systems. For these reasons, their dynamics is an active field of study \cite{Eidelstein,Nils-Oliver,Werner1,Erpenbeck,kloss2023}.
Often, the quantity of interest is the impurity Green function, and nontrivial methods have been developed, that are geared to calculating it.
This exploits the fact that a very precise description of the impurity can be achieved, employing a less precise description of the bulk.
The lack of precision where the bulk is concerned, can lead to difficulty, when attempting to study the structure of correlations that live in the bulk \cite{Gubernatis,Barzykin,Borda_2007,Lechtenberg,Florens,debertolis1}.

The fact that a Fermi liquid description applies in certain quantum impurity models, was established half a century ago, using the Numerical Renormalization Group (NRG)
\cite{CostiRMP}.
This method was able to compute the low energy many-body spectrum, whose structure
was found to be nearly identical to that of a non-interacting fermion system \cite{WilsonRMP,Nozieres1,Krishna-murthy_Wilkins_Wilson_1980}. 
The spectrum obtained for a given set of microscopic parameters
was fitted to a non-interacting fixed point Hamiltonian and leading irrelevant perturbations. 
In this way, the microscopic parameters 
could be mapped to the Fermi liquid parameters they give rise to \cite{Hewson1,Hewson2004,vondelft1,vondelft2}. 
This raises a prospect which seems unfeasible for other Fermi liquids.
Can one explicitly calculate how the quasiparticle excitations of a local Fermi liquid are constituted out of bare electronic degrees of freedom?
This question is particularly challenging to answer when a dynamic quantum impurity is imbedded in a host in which electrons experience 
 a non-trivial potential landscape, such as disorder, or electrostatic gates \cite{Dobrosavljevic_Kirkpatrick_Kotliar_1992,zarand,ALEINER,Ribhu,Ullmo2,Dong,Ullmo1,Miranda_Dias_da_Silva_Lewenkopf_2014,Slevin_Kettemann_Ohtsuki_2019}.  
Using scanning gate microscopy, experimentalists have succeeded in obtaining a real space picture of quasiparticle excitations in such systems \cite{Brun1,Brun2,Brun3}.
The microscopic modelling of these experiments may benefit from the development of new methods. One reason why NRG is inadequate, is that it 
discretizes the bulk on what is called a logarithmic energy grid. This throws away short wave-length information required to achieve 
good spatial resolution of many-body correlations in the vicinity of the impurity \cite{Barzykin,Borda_2007,Affleck}. A second reason is that NRG discards all but the very lowest
single-quasiparticle excitations before their long wave length structure is fully resolved \cite{Peters,Weichselbaum}.

Another single-particle-like picture, that is distinct from Fermi liquid theory, is important for understanding many-body
correlations in the ground states of quantum impurity problems \cite{Bravyi_Gosset_2017,Debertolis_Florens_Snyman_2021}. 
Associated with this picture are a set of bare single-particle orbitals
called natural orbitals \cite{Lowdin_1955,Davidson}. They form the single-particle basis in which a correlated ground state is expressed as a linear combination of the fewest
number of Slater determinants. Natural orbital methods have an established role in quantum chemistry, where they successfully account
for chemical properties of strongly correlated few-electron systems \cite{CASSCF_Review,Li_Paldus_2005}. 
There has recently been increased interest in the applications of
natural orbital methods in condensed matter \cite{Teemu1,vanhala}, including quantum
impurity problems \cite{Zgid_Gull_Chan_2012,He_Lu_2014}, where high accuracy in very large systems have been achieved at modest computational cost \cite{debertolis1}.
The present study was inspired by the question as to whether natural orbitals can shed light on the structure of quasiparticle excitations in a local
Fermi liquid.   

Our main results are as follows. 
We have developed methods for analyzing single-quasiparticle excitations in a large finite system consisting of a quantum impurity 
hybridized with a noninteracting host. It clarifies the relationship between the
single-particle eigenstates of the noninteracting effective quasiparticle Hamiltonian, and the associated many-body eigenstates of the original 
interacting Hamiltonian. This connection allows one to split the many-body eigenstate into two parts, corresponding respectively 
to a bare electron on top of the ground state and to the dressing of the bare electron by particle-hole pairs. 
We developed an accurate ansatz for single-quasiparticle excitations on top of the ground state 
in terms of natural orbitals. This allowed us to 
investigate quasiparticle excitations that are discarded in NRG, and to investigate the structure of the dressing of bare electrons in real space.
We could study a bulk consisting of thousands of sites and resolve the wave function with single site precision. We thus obtained
results on the non-trivial power-law decay of dressing with distance, that NRG is unable to resolve. 
Our results could lead to advances in the study of spatial features of low energy excitations in local Fermi liquids, including the modelling of realistic 
environments, such as atomic lattices, static disorder, and mesoscopic electronic devices.
It could also lead to the further refinement of natural orbital methods to study
the dynamics of quantum impurities.  

The rest of the Article is structured as follows. In Section~\ref{sec2} we introduce the Single Impurity Anderson Model \cite{Anderson}, that we will focus on.
We review elementary aspects of the local Fermi liquid theory that applies to it at low energies, 
and formulate the questions regarding its quasiparticle excitations that guided the present study.
We explain the limitations of NRG, and review the natural orbital methods
that apply to the study of the model's ground state. In Section~\ref{sec3} we present the theoretical developments that our study contribute to 
local Fermi liquid theory. In Section~\ref{sec4} we employ the theory that we developed in Section~\ref{sec3} to numerically study the 
structure of single-quasiparticle excitations. Section~\ref{sec5} contains a summary of our main results, and concluding remarks.
As evidence that our natural orbital methods are sufficiently accurate for our purposes,  in Appendix \ref{appa} we compare to results obtained by other means.

\section{Background and aims of the study}
\label{sec2}

\subsection{Local Fermi liquid theory and the Single Impurity Anderson Model}

In this Article,
our primary focus will be on the Single Impurity Anderson Model (SIAM) \cite{Anderson,hewsonbook}. It describes  $\hat d^\dagger_\sigma$ electrons in a localized orbital
interacting via Coulomb repulsion $U$ and hybridizing with the non-interacting $\hat c_{k\sigma}^\dagger$ electrons in a Fermi sea.
The Hamiltonian reads  
\begin{align}
\hat H=&U\hat n_{d\uparrow}\hat n_{d\downarrow}+\varepsilon_d\left(\hat n_{d_\uparrow}+\hat n_{d\downarrow}\right)
+V\sum_\sigma\left(\hat d^\dagger_\sigma \hat \psi_{0\sigma}+\hat \psi_{0\sigma}^\dagger \hat d_\sigma\right)\nonumber\\
&+\sum_{k=1}^L\sum_\sigma \varepsilon_k \hat c_{k\sigma}^\dagger \hat c_{k\sigma},\label{eqhamsiam}
\end{align}
where $\hat n_{d\sigma}=\hat d_\sigma^\dagger \hat d_\sigma$,
and $\hat \psi_{0\sigma}^\dagger=\frac{1}{\sqrt{L}}\sum_{k=1}^L\hat c_{k\sigma}^\dagger$
creates an electron on the lattice site closest to the impurity. To quantify the hybridization, it is conventional to cite the spectral density 
\begin{equation}
\Delta=\pi\nu V^2\label{eqhyb}
\end{equation}
at the Fermi level, of the operator $V\hat\psi_0$ in the infinite system
uncoupled from the impurity, rather than $V$. In the above expression, $\nu$ is the bulk density of states per unit volume at the Fermi level.
We measure single-particle energies such that the Fermi energy lies at zero.
If $-\varepsilon_d> \Delta$ and  $\varepsilon_d + U> \Delta$, the ground state occupation probability of the $d$-orbital is approximately one, meaning it possesses a
spin-1/2 magnetic moment. 
We will focus on the particle-hole symmetric case, where the energy of the $d$-orbital is tuned such that $\varepsilon_d=-U/2$, and the
condition for single occupation of the $d$-orbital becomes $U/2>\Delta$.
Virtual processes in which the singly-occupied impurity is emptied and then reoccupied, or doubly-occupied and then returned to single
occupancy, lead to an effective antiferromagnetic spin-exchange interaction between the impurity and the band. Kondo physics results,
associated with an emergent
Kondo scale $\tk$, given by the inverse of the $d$-orbital's susceptibility to being spin-polarized by a local magnetic field
\begin{equation}
\tk=\frac{1}{4\chi},~~~\chi=\lim_{B\to0}\left<\left(\hat n_{d\uparrow}-\hat n_{d\downarrow}\right)\right>_B/B.
\label{eqtk}
\end{equation}
Here $\left<\ldots\right>_B$ denotes the ground state expectation value when the Hamiltonian is perturbed by the local field
\begin{equation}
\hat H_B=\hat H-\frac{B}{2}\left(\hat n_{d\uparrow}-\hat n_{d\downarrow}\right).
\label{eqhb}
\end{equation}
Krishna-Murthy, Wilkins and Wilson (KWW) \cite{Krishna-murthy_Wilkins_Wilson_1980},  
showed by means of NRG that the SIAM flows to strong coupling, and reaches a stable infrared fixed point below $\tk$. 
At energies sufficiently lower than the Kondo scale, the many-body energy spectrum of the model resembles that of a non-interacting 
Fermi gas, i.e. it is given by
\begin{equation}
E(\{\nu_{\alpha\sigma}\})=\text{const.}+\sum_{\alpha}E_\alpha n_{\alpha\sigma},
\end{equation}
where the quantum numbers $n_{\alpha\sigma}\in\{0,1\}$ can be interpreted as the occupation numbers of quasiparticle orbitals.
While no explicit mapping 
relating the low-energy quasiparticles to the bare fermions appearing in (\ref{eqhamsiam}) have ever been derived,
{\it a priori} one expects the quasiparticle creation and annihilation 
operators to be nonlinear functions of the bare fermion operators $\hat c_{k\sigma}$, $\hat c_{k\sigma}^\dagger$, $\hat d_\sigma$ and $\hat d_\sigma^\dagger$.   
In other words, we expect the quasiparticles to be fermions dressed with particle-hole pairs.

Hewson \cite{Hewson1}, 
building on ideas of Nozi\`eres \cite{Nozieres1}, also introduced a quasiparticle Hamiltonian for the SIAM, as a starting point for a ``renormalized'' perturbation theory.
Hewson relied on the generic behavior of the 
(zero-temperature) retarded self-energy of a class of fermionic models that includes the SIAM, namely that the self-energy can be expanded in frequency around
$\omega=0$,
\begin{equation}
\Sigma(\omega)=\Sigma(0)+\Sigma'(0)\omega+\Sigma^\text{rem}(\omega),
\end{equation}
and that $\Sigma(0)$ and $\Sigma'(0)$ are both real. 
The interacting retarded Green function for the $d$-orbital reads $G_{\sigma\sigma'}(\omega)=\delta_{\sigma\sigma'}/(\omega-\varepsilon_d+i\Delta-\Sigma(\omega))$. 
Hewson rewrote $G_{\sigma\sigma'}(\omega)$
as
\begin{equation}
G_{\sigma\sigma'}(\omega)=\frac{z\delta_{\sigma\sigma'}}{\omega-\tilde \varepsilon_d+i\tilde \Delta-\tilde \Sigma(\omega)},\label{eqgd}
\end{equation}
where $z=1/[1-\Sigma'(0)]$, $\tilde \varepsilon_d=z[\varepsilon_d+\Sigma(0)]$ and $\tilde \Sigma(\omega)=z\Sigma^\text{rem}(\omega)\propto \omega^2$. 
(In the particle-hole symmetric case, $\tilde \varepsilon_d=0$.)
At sufficiently low frequencies/energies, $\tilde \Sigma(\omega)$ makes a very small contribution to the denominator, and can be neglected.
Hewson therefore defined a quasiparticle Hamiltonian $\hat H_\text{qp}$ that has the same form as the non-interacting version of (\ref{eqhamsiam}), with $U=0$. Furthermore
he made the replacements $V\to \tilde V=\sqrt{z} V$ and $\varepsilon_d \to \tilde \varepsilon_d$. If we denote the retarded $d$-orbital Green function
of Hewson's quasiparticle Hamiltonian by $\tilde G_{\sigma,\sigma'}(\omega)$, and drop $\tilde \Sigma(\omega)$ in the expression (\ref{eqgd}) for $G_{\sigma\sigma'}(\omega)$,
then $G_{\sigma\sigma'}(\omega)=z\tilde G_{\sigma\sigma'}(\omega)$. The elastic scattering amplitudes for a bare electrons incident on the impurity are
determined by the transfer matrix
\begin{eqnarray}
T_{kk'}(\omega)&=&\frac{1}{L}\frac{V}{\omega-\varepsilon_k}G_{\sigma\sigma}(\omega)\frac{V}{\omega-\varepsilon_{k'}}\nonumber\\
&\simeq&\frac{1}{L}\frac{\tilde V}{\omega-\varepsilon_k}\tilde G_{\sigma\sigma}(\omega)\frac{\tilde V}{\omega-\varepsilon_{k'}}.
\end{eqnarray}
The second line of the above equation implies that at low energies, 
{\it bare} electrons scatter off the impurity as if their dynamics are described by the non-interacting Hamiltonian $\hat H_\text{qp}$.  
Does this mean that our {\it a priori} expectations about KWW's quasiparticles were wrong? Could the quasiparticles be bare electrons
described by Hewson's $\hat H_\text{qp}$? The answer
is ``no''. $\hat H_\text{qp}$ has a $d$-electron 
Green function that is off by a factor $z$ compared to the original Hamiltonian (\ref{eqhamsiam}). 
Thus at least the $d$-orbital in $\hat H_\text{qp}$ is an effective, rather than a bare degree of freedom.
The fact that low energy bare electrons scatter elastically does however imply the following.
If the system is prepared in an eigenstate corresponding to a  single 
quasiparticle on top of the ground state, no particle-hole pairs escape to infinity. Instead, the particle-hole pairs that dress a bare electron to make up
a quasiparticle excitation, are confined to a finite region around the impurity. If we write the creation operator for a quasiparticle excitation, as
\begin{equation}
y_d \hat d^\dagger +\sum_{k} y_k\hat c_{k\sigma}^\dagger + \mbox{higher order terms in} 
\{\hat d^\dagger_\sigma, \hat d_\sigma, \hat c^\dagger_{k,\sigma}, \hat c_{k\sigma}\}, \label{eqqpop}
\end{equation}
then the higher order terms create particle-hole pairs in the vicinity of the impurity.
\subsection{Aims of this Article}
In this study we want to shed further light on the structure of the quasiparticles associated with the SIAM. Our aims are (1) to calculate the quasiparticle 
wave function $y_d,\,y_k,\,k=1,2,\ldots,L$ in (\ref{eqqpop}) explicitly, (2) to visualize in real space the particle-hole dressed part in the vicinity of the impurity, 
described by the higher order terms in (\ref{eqqpop}), and (3) see the onset of effects associated with the remnant self-energy $\Sigma^\text{rem}(\omega)$
at increased excitation energy. These questions require the development of new methods, as we explain next.

\subsection{Limitations of NRG}

For the purposes of numerics, one has to study a system with a finite-dimensional Hilbert space. One possibility is to study a system defined on a regular finite lattice.
However, the many-body Hilbert space becomes too large for brute-force methods already at $\sim10$ sites.
A more sophisticated option is to take an infinite system, and rediscretize it. There is considerable freedom in how to discretize, and particular choices may
be better suited to subsequent approximations than others \cite{CostiRMP}.
The general rediscretization procedure works as follows. The energy band is partitioned into intervals
$\mathcal E_k$ called energy shells. One discrete mode per shell, per channel, is retained, with associated creation operator
$\hat c_{k\sigma}^\dagger=\int_{\mathcal E_k}d\varepsilon\, \hat c_{\varepsilon\sigma}^\dagger/\sqrt{\int_{\mathcal E_k} d\varepsilon}$. 
Here $\hat c_{\varepsilon\sigma}^\dagger$ creates a particle with energy $\varepsilon$ and spin $\sigma$
(in a particular channel). Impurity problems, such as the SIAM, involve local coupling to a point impurity and scattering is therefore s-wave, so that only a single channel
is involved. Discrete mode $k$'s  energy is taken as the average energy of the interval $\mathcal E_k$.  The NRG method, to which we owe many 
results on fermionic quantum impurity problems, relies on the fact that accurate
results can be obtained for low-energy properties, from a cruder description at higher energies. It discretizes the band
``logarithmically'' so that the density of discrete states scales like $1/\left|\varepsilon_k-E_\text{F}\right|$. 
This is done by letting the width of each new interval decrease by a constant factor $1/\Lambda$ compared to the previous one,
from the band edges to the Fermi energy. 
NRG proceeds with a 
sequence of RG transformations \cite{WilsonRMP}. Step $k$ involves an approximate diagonalization that
resolves energies up to shell $k$.  Before the modes in lower energy shells are included, 
high energy states found in the current iteration are discarded, 
thus avoiding the dimension of the many-body Hilbert space becoming too large to handle. 
It turns out that the logarithmic discretization
ensures accuracy at lower energies being maintained. This accuracy extends to thermodynamic quantities such as specific heat,
and impurity spectral quantities such as the local density of states of the $d$-orbital. However, due to the decimation process that
keeps the size of the Hilbert space manageable, in practice, only the lowest few single-quasiparticle excitations are found in the end.
Furthermore, NRG inherently suffers from 
poor spatial resolution of correlations in the Fermi sea, because large shells of high-wavelength modes are crudely lumped
together into a few discrete modes \cite{Barzykin,Borda_2007}. (Often only two modes are retained to represent the top and bottom quarters of the band, i.e. $\Lambda=2$.)  
Our aim is to study the structure of single-quasiparticle modes in terms of bare electron degrees of freedom. 
This requires us to resolve spatial features all the way down to the Fermi wavelength, which NRG cannot do. 
Natural orbital methods
have recently been shown capable of the required spatial resolution, where ground state correlations are concerned \cite{debertolis1}. 
In the next subsection we review the application of natural orbital methods to impurity ground state problems.

\subsection{Covariance matrix and natural orbitals}

Given a set of fermionic creation and annihilation operators, associated with an orthonormal single-particle basis, and an arbitrary state $\ket{X}$ in the associated Fock space, 
the covariance matrix $C(\ket{X})$ is defined as 
\begin{equation}
C_{ij}\left(\ket{X}\right)=\bra{X} \hat c_i^\dagger \hat c_j \ket{X}.\label{eqcovmat}
\end{equation}
We will denote the eigenvectors and eigenvalues of the ground state covariance matrix as follows. 
\begin{equation}
\bm x_\alpha=(x_{1,\alpha},\ldots,x_{L,\alpha})^\tran,~~~C(\gsket)\bm x_\alpha=\lambda_\alpha \bm x_\alpha.\label{eqcoveig}
\end{equation}
We label eigenvectors and eigenvalues such that
$\lambda_1\geq \lambda_2\geq\ldots\geq\lambda_L$.
The eigenvalues of the covariance matrix are independent of the single-particle basis associated to $\hat c_i$. The fermionic creation operators  
\begin{equation}
\hat q_\alpha^\dagger=\sum_{i=1}^L \hat c_i^\dagger x_{i\alpha}, \label{eqnatorb}
\end{equation}
associated to eigenstates of the covariance matrix are known as the natural orbital basis \cite{Davidson}. 
This basis is independent of the the single-particle basis associated to $\hat c_i$.
The eigenvalues $\lambda_\alpha$ are the ground state occupation probabilities $\gsbra \hat q_\alpha ^\dagger \hat q_\alpha \gsket$ of the natural orbitals.
When fermions are held at a finite density, the exclusion principle forces many natural orbitals to have ground state occupation probabilities near unity.
Orbitals that are nearly filled or empty, are inert and cannot participate in many-body correlations. 
For a given small positive number $\epsilon$, this motivates us to define three sets  
\begin{eqnarray}
\mathcal O&=&\left\{\alpha|\lambda_\alpha>1-\epsilon\right\},\nonumber\\
\mathcal C&=&\left\{\alpha|\epsilon<\lambda_\alpha<1-\epsilon\right\},\nonumber\\
\mathcal U&=&\left\{\alpha|\lambda_\alpha<\epsilon\right\},\label{eqsectors}
\end{eqnarray}
which we call the occupied, correlated, and unoccupied sectors respectively. (Quantum chemists refer to these sets as the inactive, active and virtual spaces.)

In a quantum impurity problem, the bulk remains non-interacting, and the correlated sector only contains a vanishing fraction of the total number
of orbitals in a large system. It was recently realized that for a generic fermionic impurity problem, the number of orbitals in the correlated sector 
is proportional to $-{\rm ln}\,\epsilon$ at sufficiently small $\epsilon$, with a proportionality constant that remains finite in the thermodynamic limit. 
As a consequence, the ground state of a generic fermionic impurity model can be approximated as follows, if the natural orbitals are known \cite{Debertolis_Florens_Snyman_2021}.
Consider a model with $N$ particles. At given $\epsilon$, let $N_{\mathcal O}(\epsilon)$ and $N_{\mathcal C}(\epsilon)$ be the numbers of orbitals in respectively 
the occupied and correlated sectors. For sufficiently small $\epsilon$, we can approximate
the orbitals in the occupied sector as fully occupied, and those in the unoccupied sector as completely empty, meaning that in the ground state the correlated sector 
contains $m=N-N_{\mathcal C}(\epsilon)$ particles distributed among $N_{\mathcal C}(\epsilon)$ orbitals. Let
\begin{equation}
\fermi=\hat q_{N_{\mathcal O}}^\dagger\ldots \hat q_1^\dagger\ket{0},\label{eqfermi}
\end{equation}
be the Fermi sea corresponding to the completely filled occupied sector. Given a set of $n$ 
orbitals $\sset{\alpha}{n}\subset\mathcal C$ belonging to the correlated sector, we define    
\begin{equation}
\ket{\sset{\alpha}{n}}=\left({\prod}'_{\alpha\in\sset{\alpha}{n}}\hat q_\alpha^\dagger\right)\fermi.\label{eqcorbasis}
\end{equation}
Here the prime denotes a fixed ordering of operators (say $\alpha$ decreasing from left to right), to remove ambiguity about the phase of the state.
We take as ground state ansatz, an arbitrary linear combination of $N$-particle states of the form (\ref{eqcorbasis}):
\begin{equation}
\gsket\simeq\sum_\sset{\alpha}{m}v_{\sset{\alpha}{m},\text{GS}}\ket{\sset{\alpha}{m}}.\label{eqgsansatz}
\end{equation}
The optimal state is found by minimizing the expectation value of the energy over the coefficients $v_{\sset{\alpha}{m},\text{GS}}$. 
Thus it is found that the optimal expansion coefficients $v_{\sset{\alpha}{m},\text{GS}}$ correspond to the ground state eigenvector of the effective
few-body Hamiltonian 
\begin{equation}
\left[H_\text{few\,body}\right]_{\sset{\alpha}{m},\sset{\beta}{m}}=\bra{\sset{\alpha}{m}}\hat H\ket{\sset{\beta}{m}}.
\label{eqhamfew}
\end{equation}
This type of approximation has a long history in quantum chemistry, where it is called the Complete Active Space (CAS) approach \cite{CASSCF_Review}. 
In the context of quantum impurity problems,
stronger results regarding accuracy apply than in quantum chemistry, thanks to the proven scaling of the size of the correlated sector with $\epsilon$:
The dimension of the effective few-body Hamiltonian is $\left(\begin{array}{c}N_{\mathcal C}(\epsilon)\\m\end{array}\right)$, with $m\sim N_{\mathcal C}(\epsilon)/2$,
which scales exponentially with $N_{\mathcal C}$. However, $N_{\mathcal C}$ only scales logarithmically with $\epsilon$, and the dimension of the 
Hamiltonian that has to be diagonalized therefore scales polynomially with $1/\epsilon$. These features of fermionic quantum impurity
problems have been exploited to prove that the computational complexity of finding the ground state of a fermionic quantum impurity problem 
scales quasi-polynomially with the 
inverse of the required accuracy and polynomially with the system size \cite{Bravyi_Gosset_2017}. 
What makes the proof non-trivial is the fact that the natural orbitals are defined relative to the ground state,
and therefore not known beforehand. In practice, iterative algorithms are found to work well. 
In these algorithms, a guess for the natural orbitals is recursively improved from the previous iteration's result for the approximate ground state.
This is known as the Recursive Generation of Natural Orbitals (RGNO) \cite{Li_Paldus_2005}. An early application of the method in a Condensed Matter
context considered a multi-impurity system \cite{He_Lu_2014}. 
Below we find that for the SIAM, the ground state energy can be determined to an accuracy of 5\% of the Kondo temperature, for realistic
Kondo temperatures of 1 \% of the band width, and well-developed correlations (quasiparticle weight $z=0.2$), 
using very modest resources, namely a correlated sector consisting of 12 orbitals (six spin up and six spin down),
containing 6 particles (three spin up and three spin down).  

\subsection{Discretizations employed}

Below, we will study quasiparticles using both NRG and natural orbital methods. As stated above, NRG requires a logarithmic discretization.
Natural orbital methods allow more freedom. Here we provide details regarding the different discretizations we will employ.
For simplicity, we will work with a continuum model that has a 
half-bandwidth $D$ throughout, and assume that the continuum model has a flat density of states per unit length so that
\begin{equation}
\Delta=\frac{\pi V^2}{2D}.\label{eqdelta2}
\end{equation} 
For a logarithmic discretization, the conduction band is divided into intervals 
\begin{equation}
\left[-D\Lambda^{-k},-D\Lambda^{-k-1}\right)\text{ and }\left(D\Lambda^{-k-1},D\Lambda^{-k}\right],
\label{eqshells}
\end{equation}
with $\,k=0,1,2,\ldots$.
The energies of the rediscretized conduction band orbitals are
\begin{equation}
\varepsilon_{k\pm}=\pm \frac{D}{2\Lambda^k}\left(\frac{1}{\Lambda} +1\right).
\end{equation}
After the above discretization, the discrete system still possesses an infinite number of modes. NRG proceeds by 
turning the diagonal kinetic term into a tri-diagonal form  
\begin{align}
\hat H=&U\hat n_{d\uparrow}\hat n_{d\downarrow}+\varepsilon_d\left(\hat n_{d_\uparrow}+\hat n_{d\downarrow}\right)
+V\sum_\sigma\left(\hat d^\dagger_\sigma \hat \psi_{0\sigma}+\hat \psi_{0\sigma}^\dagger \hat d_\sigma\right)\nonumber\\
&+\sum_{j=0}^\infty\sum_\sigma t_j  \left(\hat \psi_{j\sigma}^\dagger \hat  \psi_{j+1\sigma}+\hat \psi_{j+1\sigma}^\dagger \hat  \psi_{j\sigma}\right),\label{eqchain}
\end{align}
called the Wilson chain, with the off-diagonal elements $\propto t_j$ describing hopping along the chain.
Their explict form can be looked up in \cite{CostiRMP}. The first chain site corresponds to the lattice site $0$ that is  
directly coupled to the impurity. Hopping amplitudes decrease like $\Lambda^{-j/2}$ along the chain, with site $j$ associated with energy scale $D\Lambda^{-j/2}$.
 An infrared cut-off is imposed by truncating the chain after $L$ sites, which yields a finite system, and a smallest energy scale $\sim D\Lambda^{-L/2}$. 
 We will perform both NRG and natural orbital calculations on the Wilson chain.
 
The Wilson chain is not suitable for resolving the real space structure of quasiparticle excitations.
We will therefore also use a second, more suitable discretization, in conjunction with natural orbital methods.
For this purpose, we take the discrete energies of the finite system to be
\begin{equation}
\varepsilon_k=\frac{k-\Omega-1}{\Omega}D,~~~L=2\Omega+1.\label{eqlindiscr}
\end{equation}
To study spatial structure, we interpret the $\hat c_{k\sigma}^\dagger$ operators associated with these energies
as the even modes of a one-dimensional lattice with $4\Omega+1$ sites, the central site ($j=0$) of which is side-coupled to the 
$d$-orbital. We introduce symmetrized position representation operators 
\begin{equation}
\hat \gamma_{j\sigma}=\sqrt{\frac{1}{2\Omega+1}}\sum_{k=1}^{2\Omega+1}
\cos\left[\frac{\pi j (2k-1)}{4\Omega+2}\right]\hat c_{k\sigma}, \label{eqsymfieldop}
\end{equation}
for $j=0,\ldots,2\Omega$. (Symmetrization means that $\hat \gamma_{j\sigma}$ is one half times the sum of the operators that respectively annihilate an electron 
on sites $j$ and $-j$. This way we avoid having to introduce creation operators for the odd parity modes that do not couple to the $d$-orbital.)
The $\hat\gamma_{j\sigma}$ operators obey
\begin{equation}
\left\{\hat\gamma_{j,\sigma},\hat\gamma_{j',\sigma'}^\dagger\right\}=\frac{\delta_{j,j'}\delta_{\sigma,\sigma'}}{(2-\delta_{j,0})}.
\label{gammanorm}
\end{equation}
Since the infrared energy cutoff $D/\Omega$ scales inversely rather than exponentially with system size, very large systems are required to resolve the
emergent infrared physics of quantum impurity models. Natural orbital methods have proved capable of this task, where ground state properties are
concerned.

\section{Theoretical developments}
\label{sec3}
In this section, we present two theoretical developments, that we made, and which allows us to learn more about the the structure of
single-quasiparticles in a local Fermi liquid, than was known before. The first is a wave function picture of local Fermi liquid theory, applicable to finite systems. This
complements Hewson's Green function picture for infinite systems. It provides us with tools to analize the structure of single-quasiparticle excitations.
The second development is an ansatz for single-quasiparticle excitations in terms of natural orbitals.
We will find that results are accurate in a regime where correlations are sufficiently strong to study non-trivial local Fermi liquids, 
and amenable to discretization on a large regular energy grid.

\subsection{Wave function picture of local Fermi liquids}
\label{wavepic}
Suppose the ground state $\gsket$, as well as excited states $\ket{\text{p},n\sigma}$ with a single quasiparticle on top of the ground state,
are known to good accuracy. How do we find the (dominant) linear part, cf.~(\ref{eqqpop}) of a quasiparticle operator $\hat a_{n\sigma}$
such that 
\begin{equation}
\ket{\text{p},n\sigma}{\simeq} \hat a_{n\sigma}^\dagger \gsket? \label{eqqphyp}
\end{equation}
We can answer this question by setting
\begin{equation}
\hat a_{n\sigma}=y_{dn} \hat d_\sigma + \sum_{k=1}^L y_{kn}\hat c_{k\sigma},~~~|y_{dn}|^2+\sum_{k=1}^L |y_{kn}|^2=1,\label{eqyconstraint}
\end{equation}
and maximizing the object function
\begin{equation}
p_n=\left|\gsbra \hat a_{n\sigma}\ket{\text{p},n}\right|^2,\label{eqw1}
\end{equation}
over all unit-length vectors
\begin{equation}
\bm y_n=(y_{dn}, y_{1n},\ldots, y_{Ln})^\tran.\label{eqyvec}
\end{equation}
The optimal solution is found to be
\begin{equation}
y_{dn} =\frac{ \bra{\text{p},n} \hat d^\dagger_{\sigma}\gsket}{\sqrt{p_n}},~~~
y_{kn} =\frac{ \bra{\text{p},n} \hat c_{k\sigma}^\dagger\gsket}{\sqrt{p_n}},
\label{eqtildey}
\end{equation} 
The optimal $p_n$ that results is
\begin{equation}
p_n=\left|\gsbra \hat d_{\sigma}\ket{\text{p},n}\right|^2+\sum_{k=1}^L\left|\gsbra \hat c_{k\sigma}\ket{\text{p},n}\right|^2.\label{eqw2}
\end{equation}
The quantity $p_n$ is the probability to measure only a single bare electron excitation on top of the ground state when the state $\ket{\text{p},n\sigma}$ is prepared.
The quantity
\begin{equation}
\delta\rho(E) =\sum_n(1-p_n)\delta(E-E_n),\label{eqdrho}
\end{equation}
equals the difference in density of single-particle states between a system of bare non-interacting electrons with the same spectrum as the quasiparticles, 
and the actual system. (Here we do not mean ``density of states per unit volume'', but the actual denisty of states, that diverges in the thermodynamic limit.)
The density of states difference 
remains finite in the thermodynamic limit, and quantifies the extent to which bare electrons are dressed in order to form quasiparticles.
It is tempting to 
use Fermi liquid green functions, with the remnant self-energy $\Sigma^\text{rem}(\omega)$ neglected, to evaluate 
$\delta\rho$ at low energies in the thermodynamic limit. 
Since the quasiparticle Green function and the actual single-electron Green function only differ on the 
$d$-orbital, one finds $\delta\rho(\omega)=\text{Im}\left[\tilde G_{\sigma\sigma}(\omega)-G_{\sigma\sigma}(\omega)\right]/\pi$. At the Fermi level,
where $\Sigma^\text{rem}(\omega)=0$, this is certainly valid, and gives
\begin{equation}
\delta\rho(0)=\frac{1}{\pi}\left(\frac{1}{\tilde\Delta}-\frac{1}{\Delta}\right)=\frac{1}{\pi\Delta}\left(\frac{1}{z}-1\right).\label{eqdrhoz}
\end{equation}
In a large but finite system, in which the spacing of single-particle levels near the Fermi energy is $\delta E$, this gives
\begin{equation}
z\simeq \frac{1}{1+p_n\pi\Delta/ \delta E}.\label{eqzitop}
\end{equation}
Thus the wave function picture of a local Fermi liquid furnishes us with an 
interpretation of the quasiparticle weight in terms of the bare electron occupation probability. 

Above the Fermi level, the reasoning that led to (\ref{eqdrhoz}) 
would predict that $\delta\rho(\omega)$ decreases proportional to $1/(\omega^2+\tilde\Delta^2)$,
whereas we expect $p_n$ to decrease and hence $\delta\rho$ to increase as the quality of 
quasiparticles deteriorates with increasing excitation energy. Below we perform numerics to resolve the behaviour of $\delta\rho(\omega)$ at $\omega>0$.

A mode-resolved measure $D_{kn}$ of how a bare electron is dressed to form the excitation $\ket{\text{p},n\sigma}$ is obtained as follows. 
Since $\ket{\text{p},n\sigma}=\sqrt{p_n}\hat a_{n\sigma}^\dagger\gsket+\ket{\mbox{dressing}}$, we set
\begin{align}
D_{kn}&\equiv\bra{\mbox{dressing}}\hat c_{k\sigma}^\dagger\gsket\nonumber\\
&=\bra{\text{p},n\sigma}\hat c_{k\sigma}^\dagger\gsket-\sqrt{p_n}\gsbra \hat a_{n\sigma}\hat c_{k\sigma}^\dagger\gsket\nonumber\\
&=\bra{\text{p},n\sigma}\hat c_{k\sigma}^\dagger\gsket-\sqrt{p_n}\gsbra \{\hat a_{n\sigma},\hat c_{k\sigma}^\dagger\}\gsket\nonumber\\
&~+\sqrt{p_n}\gsbra \hat c_{k\sigma}^\dagger \hat a_{n\sigma} \gsket\nonumber\\
&=\underbrace{\bra{\text{p},n\sigma}\hat c_{k\sigma}^\dagger\gsket-\sqrt{p_n}y_{kn}}_{=0}+\sqrt{p_n}\gsbra \hat c_{k\sigma}^\dagger \hat a_{n\sigma} \gsket\nonumber\\
&=\left[\sqrt{p_n}C( \gsket)\bm y_n\right]_k,\label{eqdress}
\end{align}
where we used the definition of $y_{kn}$ (\ref{eqtildey}) twice in the last two lines. This measure equals zero when $\ket{\text{p},n\sigma}$ is associated with a bare
electron on top of the ground state (i.e. $p_n=1$). When the index $k$ refers to wave number, a spatially resolved 
measure of dressing is obtained by Fourier transforming to real space. 
Given the arguments presented above, we expect to see a signal that decays to zero as we move away from the
impurity.

Intuitively, we suspect a close link between $\bm y_n$ and the eigenvectors of the type of single-quasiparticle Hamiltonian identified by Hewson. 
To explore this link, consider the quantities $\bra{\text{p},n\sigma} [\hat H,\hat d^\dagger_{\sigma}]\gsket$ and 
$ \bra{\text{p},n\sigma} [\hat H,\hat c_{k\sigma}^\dagger]\gsket$, where $\hat H$ is the particle-hole symmetric SIAM.
This leads to the equations
\begin{align}
(E_n-E_\text{GS})&\bra{\text{p},n\sigma} \hat c_{k\sigma}^\dagger\gsket\nonumber\\
&= \frac{V}{\sqrt{L}}\bra{\text{p},n\sigma} \hat d^\dagger_{\sigma}\gsket+\varepsilon_k \bra{\text{p},n} 
\hat c_{k\sigma}^\dagger\gsket,\label{eqqpwav1}
\end{align}
and
\begin{align}
(E_n-E_\text{GS})&\bra{\text{p},n\sigma} \hat d_{\sigma}^\dagger\gsket\nonumber\\
&= U\bra{\text{p},n\sigma} (\hat n_{d-\sigma}-1/2)\hat d^\dagger_{\sigma}\gsket\nonumber\\
&~~+ \frac{V}{\sqrt{L}}\sum_{k=1}^L\bra{\text{p},n\sigma} 
\hat c_{k\sigma}^\dagger\gsket.\label{eqqpwav2}
\end{align}
If
\begin{equation}
\frac{U\bra{\text{p},n\sigma} (\hat n_{d-\sigma}-1/2)\hat d^\dagger_{\sigma}\gsket}
{\frac{V}{\sqrt{L}}\sum_{k=1}^L\bra{\text{p},n\sigma} 
\hat c_{k\sigma}^\dagger\gsket}\equiv(1-z)\label{eqdefz1}
\end{equation}
does not vary much as a function of $n$, we can
employ it as a definition of the quasiparticle weight $z$.
That the above-defined $z$ is approximately independent of $n$ for low energy states can be made plausible by noting that
$(\hat n_{d-\sigma}-1/2)\hat d^\dagger_{\sigma}$ and $\sum_{k=1}^L\hat c_{k\sigma}^\dagger/\sqrt{L}$ probe the system locally at one end,
while $n\to n+1$ affects the phase difference between amplitudes on adjacent sites by an amount $\sim 1/L\ll1$ only.
Note that via (\ref{eqqpwav2}), (\ref{eqdefz1}) is equivalent to
\begin{equation}
z=\frac{(E_n-E_\text{GS})\bra{\text{p},n\sigma} \hat d_{\sigma}^\dagger\gsket}{ \frac{V}{\sqrt{L}}\sum_{k=1}^L\bra{\text{p},n\sigma}\hat c_{k\sigma}^\dagger\gsket}.\label{eqdefz2}
\end{equation} 
We substitute (\ref{eqdefz1}) into (\ref{eqqpwav2}) and define 
\begin{equation}
\psi_{dn} =\frac{ \bra{\text{p},n\sigma} \hat d^\dagger_{\sigma}\gsket}{\sqrt{z}},~~~
\psi_{kn} =\bra{\text{p},n\sigma} \hat c_{k\sigma}^\dagger\gsket,
\label{eqpsi}
\end{equation} 
and  $\tilde V=\sqrt{z} V$, to obtain the effective single-quasiparticle Schr\"odinger equation
\begin{eqnarray}
(E_n-E_\text{GS})\psi_{kn}&=& \frac{\tilde V}{\sqrt{L}}\psi_{dn}+\varepsilon_k \psi_{kn} ,\nonumber\\
(E_n-E_\text{GS})\psi_{dn}&=& \frac{\tilde V}{\sqrt{L}}\sum_{k=1}^L\psi_{kn},\label{eqqpschrod}
\end{eqnarray}
associated with a non-interacting SIAM ($U=0$) and a renormalized hybridization $\tilde \Delta= z \Delta$. 
This is precisely the form identified by Hewson, and implies that $\psi_{dn}=\sqrt{p_n/z}\,y_{dn}$, $\psi_{kn}=\sqrt{p_n}\, y_{kn}$.
   
\subsection{Ansatz}
\label{ansatz1}
Next, we explain our approach to generalizing natural orbital methods to excited states.   
We consider $N\pm1$ particle excited states that are obtained by adding or removing
a particle from the non-interacting system, and adiabatically switching on the interaction. If the low-energy physics is that of a local Fermi liquid,
these excitations will be single-quasiparticle- or -hole-like in nature.

Based on the notion that the natural orbital basis minimizes the Hamiltonian's ability to generate particle-hole pairs, we 
make the following ansatz for excited states consisting of one particle on top of the ground state.
\begin{equation}
\ket{\text{p},k}_1=\sum_{\alpha\in\mathcal U}u_{\alpha,k}\,\hat q_\alpha^\dagger\gsket
+\sum_\sset{\alpha}{m+1}v_{\sset{\alpha}{m+1},k}\ket{\sset{\alpha}{m+1}}.\label{eqansatz1p}
\end{equation}
This amounts to assuming that
\begin{enumerate}
\item adding a particle leaves the occupied sector undisturbed,
\item there is at most one particle in the unoccupied sector,
\item and if the added particle ends up in the unoccupied sector, the particles in the correlated sector are are not disturbed from the way they were configured in the ground state.
\end{enumerate}
(Of these assumptions, the last one (3) seems the most arbitrary. In Appendix~\ref{appa} we therefore investigate the consequences of 
not making this assumption. The conclusion is that assumption (3) does in fact apply in the local Fermi liquid regime.)  
We find the coefficients $u_{\alpha,k}$ and $v_{\sset{\alpha}{m+1},k}$ variationally. Formally, we first find the lowest energy state of this form,
then we vary orthogonal to the lowest energy state, to get the second lowest energy state and so on. 
Luckily, one does not ever explicitly have to parametrize the space orthogonal to the states found already:
it is easy to show that  the coefficients of the states obtained in the procedure are simply the eigenvectors of the effective Hamiltonian 
\begin{equation}
H_\text{p}=\left(\begin{array}{cc}H_{\mathcal U\mathcal U} & H_{\mathcal U\mathcal C}\\
H_{\mathcal U\mathcal C}^\dagger&H_{\mathcal C\mathcal C}
\end{array}
\right),
\label{eqhammat1}
\end{equation}
consisting of the blocks
\begin{eqnarray}
\left[H_{\mathcal U \mathcal U}\right]_{\alpha,\beta}&=&\gsbra \hat q_\alpha \hat H \hat q^\dagger_\beta
\gsket,\nonumber\\
\left[H_{\mathcal U \mathcal C}\right]_{\alpha,\sset{\beta}{m+1}}&=&\gsbra \hat q_\alpha \hat H 
\ket{\sset{\beta}{m+1}},\nonumber\\
\left[H_{\mathcal C \mathcal C}\right]_{\sset{\alpha}{m+1},\sset{\beta}{m+1}}&=&
\bra{\sset{\alpha}{m+1}} \hat H \ket{\sset{\beta}{m+1}}.\label{eqhamblocks1}
\end{eqnarray}
The dimension of this Hamiltonian is $N_{\mathcal U}+\left(\begin{array}{c}N_{\mathcal C}\\ m+1\end{array}\right)$, where $N_{\mathcal U}$ is the number of orbitals in the unoccupied sector.
Similar to the ansatz for particle-like excitations, we make an ansatz for hole-like excitations.
\begin{equation}
\ket{\text{h},k}_1=\sum_{\alpha\in\mathcal O}u_{\alpha,k}\,\hat q_\alpha\gsket
+\sum_\sset{\alpha}{m-1}v_{\sset{\alpha}{m-1},k}\ket{\sset{\alpha}{m-1}}.\label{eqansatz1h}
\end{equation}
We note that variational trial states with limited numbers of particles in the unoccupied sector and holes in the occupied sector have been
employed before in the context of quantum impurity problems \cite{Zgid_Gull_Chan_2012,Lin_Demkov_2013}, but as far as we know, not to calculate excited quasiparticle states.
      
To gauge the accuracy of the ansatz, we benchmarked it against NRG results. Details can be found in Appendix~\ref{appa}. This allowed us to identify SIAM parameters
that produce a well-developed Kondo regime, and in which the ansatz gives accurate results at an affordable computational cost.

\section{Numerical Results}
\label{sec4}

We now present numerical results that we obtained by applying the tools developed in Section~\ref{wavepic} to eigenstates of the SIAM that comprise a single quasiparticle
on top of the ground state. We obtained the latter by means of NRG, or more approximately, using the ansatz (\ref{eqansatz1p}).

\subsection{The bare electron occupation probability $p_n$ and the DOS difference $\delta\rho$.}

\begin{figure*}[htbp]
\begin{center}
\includegraphics[width=1.03\columnwidth]{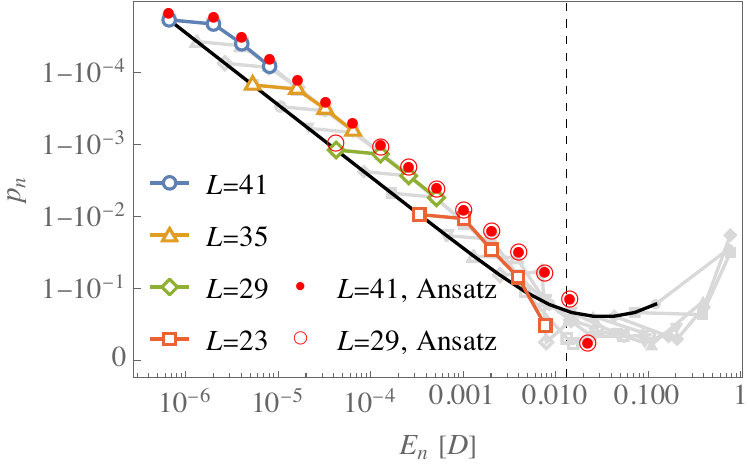}
\includegraphics[width=.97\columnwidth]{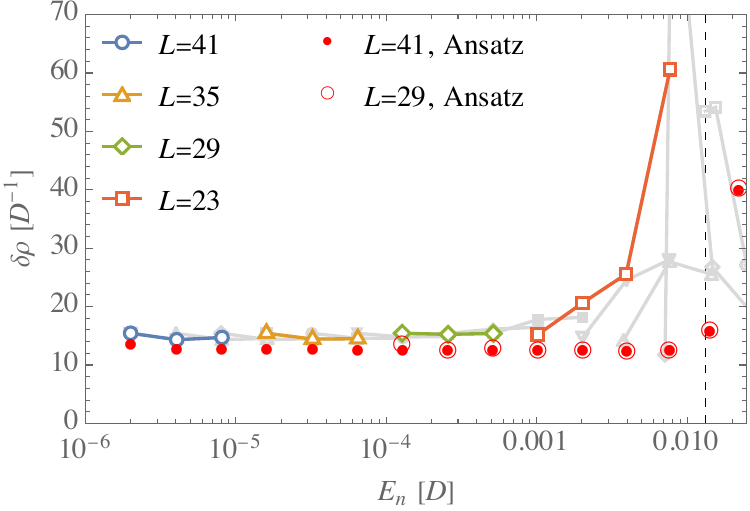}
\caption{{\bf Left panel:} The probaility $p_n$ (\ref{eqw1}) to detect only a bare electron above the ground state for the symmetric SIAM with  $\Delta=0.09 D$ and $U=0.6 D$. {\bf Right panel:} The difference $\delta\rho$ (\ref{eqdrho}) between the density of states of the single-quasiparticle 
Hamiltonian (\ref{eqqpschrod}) and the single-particle density of states of the actual Hamiltonian, calculated from the same data. Results are shown for all states with $p_n>0.5$.
The grey data represent NRG results for Wilson chains of lengths of $L=2\Omega+1$ with $\Omega=1,8,\ldots,20$ and $\Lambda=2$.
The black solid curve in the left panel connects the data points corresponding to the lowest energy states with one particle on top of the ground state, at different $L$. 
In the right panel, the states connected by this black line were omitted.
The dashed vertical line
indicates the Kondo temperature (\ref{eqtk}). NRG results at representative $L$ are highlighted as indicated in the legend. Results obtained with the Ansatz, on the same Wilson chain as used in NRG, are plotted with the symbols indicated in the legend.}
\label{fig2}
\end{center}
\end{figure*}

\begin{figure}[htbp]
\begin{center}
\includegraphics[width=1.0\columnwidth]{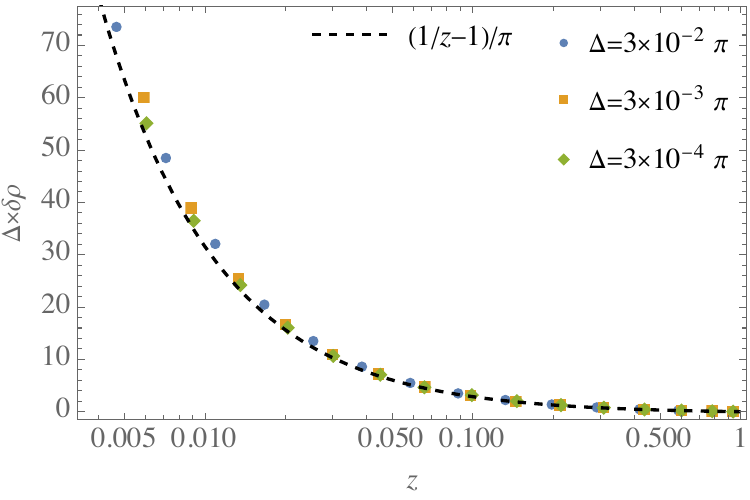}
\caption{Hybridization $\Delta$ times the density of states difference $\delta \rho$ (\ref{eqdrho}), for different $U$ and three values of $\Delta$, versus the corresponding quasiparticle weight $z$.
A dashed line indicates the estimate (\ref{eqdrhoz}).}
\label{fig2prime}
\end{center}
\end{figure}

We have performed numerical investigations on the particle-hole symmetric SIAM using NRG and the ansatz (\ref{eqansatz1p}). In this subsection we show results for $\Delta=0.09 D$ and $U=0.6 D$, which correspond to
$\tk=0.013D$.  In Figure~\ref{fig2} we present results for the bare electron occupation probability
$p_n$ (\ref{eqw1}) and the difference in single-particle DOS (\ref{eqdrho}) between the quasiparticle Hamiltonian (\ref{eqqpschrod}) and the actual system. We present 
NRG results for different Wilson chain lengths $L$ (infrared energy cut-offs) for the same SIAM parameters. 
For each $L$, NRG could find the lowest $n_L$ single-particle excitations, with $n_L$ between 2 and 4. 
(Higher excited single-quasiparticle excitations get decimated during the renormalization flow.) In order to access higher energy single-quasiparticle excitations, we
employed the ansatz (\ref{eqansatz1p}) on the same Wilson chain Hamiltonian as NRG. We used a correlated sector with six spin up and six spin down orbitals, that is half-filled for the ground state. At each $L$ we see that there is a slight decrease in $p_n$ going from the lowest single-quasiparticle excitation to the second lowest. There is a steeper decrease of $p_n$ going from the second to the third, third to fourth excitation, and so on. The fact that the lowest single-quasiparticle excitation behaves differently than the rest
has to do with the energy discretization employed by NRG, in which the inverse of the level spacing between single-particle states of the Wilson chain has a similar behaviour.
At fixed $L$, $1-p_n$ is proportional to $E_n$, for $n=2,3,4$, with an $L$-independent proportionality constant. 
For the lowest single-quasiparticle state ($n=1$) at different $L$, we also find $1-p_1\propto E_1$. We see good agreement between
NRG and the ansatz. When $L$ is varied, the $1-p_n\propto E_n$
behaviour breaks down when the infrared cut-off scale is $\sim 0.1\tk$. On the other hand, when $L$ is held fixed such that the corresponding
infrared cutoff scale is much smaller than $\tk$, the $1-p_n\propto E_n$ behaviour persists up to nearly $\tk$. We will explore the reason behind these contrasting behaviours 
further below in Figure~\ref{fig4}. 

We obtain the DOS difference $\delta \rho$
by multiplying $1-p_n$ by the density of single particle energies of the Wilson chain $=1/\text{ln}(\Lambda)E_n$, cf. (\ref{eqdrho}). To avoid having to deal with the irregularity 
of the level spacing at the lowest single-quasiparticle state, we exclude it from the presented results in the right panel of Figure~\ref{fig2}. At $E_n\ll \tk$, $\delta\rho$ is constant.
NRG is sufficient to show that $\delta\rho$ increases above the low energy plateau, if the system's infrared cut-off energy scale is larger than $0.1\tk$. However,
the ansatz is required to study single-quasiparticle excitations close to $\tk$ when the system's infrared cut-off scale is much lower than $\tk$. 
The ansatz indicates that in this case, $\delta\rho$ remains constant up to nearly $\tk$, after which it increases. An increase of $\delta\rho$, rather than a decrease, 
validates the intuition articulated in Section~\ref{sec3},
that $\delta\rho$ reveals deviations from perfect Fermi liquid behaviour when the system is probed at energies approaching $\tk$.

In Section~\ref{sec3}, we predicted (\ref{eqdrhoz}) that the value of $\delta\rho$ on the low energy plateau should be $(1/z-1)/\pi\Delta$ where $z$ is the quasiparticle weight. We test this prediction as follows. Using NRG, we calculate the plateau value of $\delta\rho$ for different $U$ and $\Delta$. For each calculation, we also determine $z$ by fitting the low energy spectrum
to that of a non-interacting SIAM ($U=0$), with renormalized $\tilde V=\sqrt{z} V$ (\ref{eqqpschrod}). The results of this analysis are shown in Figure~\ref{fig2prime}. We see very
good agreement between NRG results and our prediction. This confirms the validity of the interpretation given to the quasiparticle weight $z$ within the wave function picture
of local Fermi liquid theory, namely that it is connected to the bare electron occupation probability via (\ref{eqzitop}).     
    
\begin{figure*}[phtb]
\begin{center}
\includegraphics[width=.95\columnwidth]{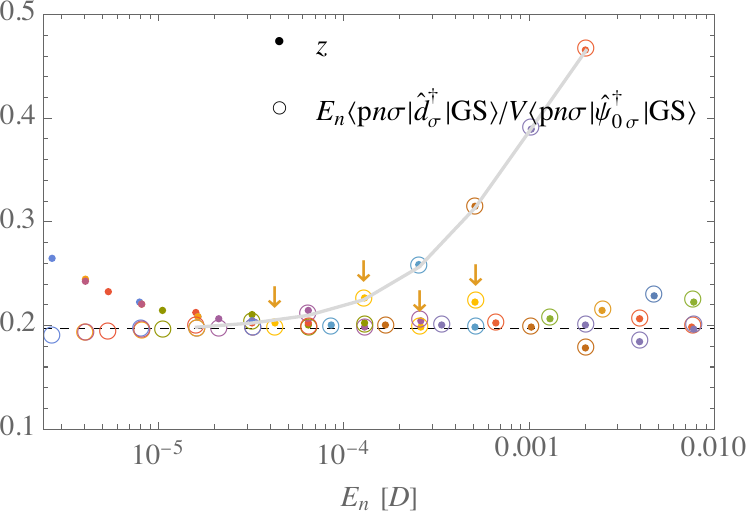}
\includegraphics[width=1.05\columnwidth]{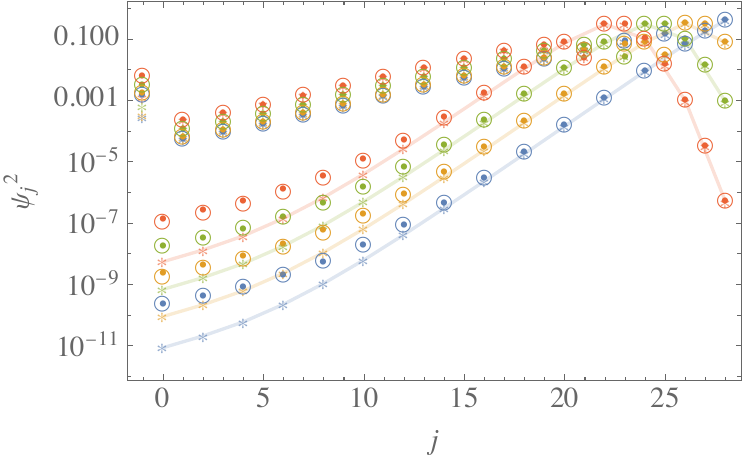}
\caption{{\bf Left panel:} The quasi-particle weight $z$, for the symmetric SIAM with  $\Delta=0.09 D$ and $U=0.6 D$. Results were obtained with NRG on Wilson chains of
lengths of $L=2\Omega+1$ with $\Omega=7,8,\ldots,20$ and $\Lambda=2$. 
Different colors represent different $L$.
Each plot marker represents a single-quasiparticle excitation. 
Dots represent the estimate (\ref{eqdefz1}) while
open disks represent the estimate (\ref{eqdefz2}). The dashed line was obtained from $z=\tilde V^2/V^2$, where $\tilde V$ was determined by fitting the 
spectrum of the quasiparticle Hamiltonian (\ref{eqqpschrod}) to the single-particle spectrum found in NRG, at $L=29$.
{\bf Rigth panel:} Square of the quasiparticle wave functions $\psi_j$, with unit-normalization, for the symmetric SIAM with  $\Delta=0.09 D$ and $U=0.6 D$ on a Wilson chain of
length of $L=29$. (Here $j=-1$ refers to the $d$-orbital.) The lowest four single-quasiparticle wave functions are shown. The dots were calculated with NRG. The open circles represent the wave functions calculated
using the single-quasiparticle Hamiltonian (\ref{eqqpschrod}). The quasiparticle weight represented by the dashed line in the left panel was employed. The stars show the 
single-particle wave functions of the non-interacting $\Delta=0.09 D$, $U=0$ problem, for comparison.}
\label{fig1}
\end{center}
\end{figure*}

\subsection{Quasiparticle weight and wave function}
\label{ssa}

In Section \ref{sec3}, we made a connection (\ref{eqpsi}) between the matrix elements $\bra{\text{p},n\sigma}\hat d_\sigma^\dagger\gsket$ and 
$\bra{\text{p},n\sigma}\hat c_{k\sigma}^\dagger\gsket$ on the one hand, and eigenvectors of a single-quasiparticle Hamiltonian (\ref{eqqpschrod}) on the other.
This rests on the assumption that a quasiparticle weight $z$, approximately independent of excitation energy,   can be defined through either 
(\ref{eqdefz1}) or equivalently (\ref{eqdefz2}). In Figure~\ref{fig1} we validate this assumption.
In the left panel, we compare several quantities calculated with NRG, for the particle-hole symmetric SIAM.
The dashed line was obtained by fitting the spectrum of single-quasiparticle Hamiltonian
(\ref{eqqpschrod}) for given quasiparticle weight $z$, to the low energy spectrum obtained by NRG. 
(This is the same procedure as was followed to obtain $z$ in Figure~\ref{fig2prime}.) The symbols represent the two equivalent expressions  
(\ref{eqdefz1}) and (\ref{eqdefz2}) evaluated for every low energy eigenstate found by NRG, that has $p_n>0.5$. Results
for Wilson chains of different lengths $L$ are shown together.  
There is a clear agreement between the conventional definition of the quasiparticle weight, represented by the dashed line, and the symbols representing (\ref{eqdefz1}) and (\ref{eqdefz2}). However, there also appear some outliers. 
Given any exact eigenstate, the quantity resperesnted by the dots should equal the quantity represented by the disks, regardless of whether they equal the 
energy-independent quasiparticle weight.
At the lowest energies (corresponding to the longest Wilson chains) the dots 
and the open disks do not lie on top of each other. This indicates that these low energy deviations
are an artefact due to the inherent numerical instability of calculating matrix elements of irrelevant operators near the infrared fixed point. 
Another branch of outliers are indicated by a solid grey curve. Tracing this curve from right to left, it eventually merges with the correct value of $z$. 
We have identified that this branch is associated with hybridization on the Wilson chain between the second single-quasiparticle eigenstate
and the lowest excited state containing two quasiparticles and one quasihole. During the renormalization flow, these two many-body levels approach each other. Initially
there is some level repulsion and associated hybridization. However, eventually (for sufficiently long Wilson chains) the two become degenerate, so that a well-developed
single-quasiparticle excitation can be distinguished. This is the point where the branch merges with the plateau indicated by the dashed line. This branch therefore
does not represent a failure of the theory developed in Section~\ref{sec3}, but rather a genuine interaction between quasiparticles at intermediate energies. 
In other words, if the finite Wilson chain with $L\leq 31$ could be weakly connected to leads, the considered outliers would represent a scattering resonance, due to an accidental degeneracy, in which an incoming electron really does scatter into two electrons and a hole. 

In the right
panel of Figure~\ref{fig1}, we plot the square of $\psi_{dn}=\bra{\text{p},n\sigma}\hat d^\dagger_\sigma\gsket/\sqrt{z}$ and 
$\psi_{jn}=\bra{\text{p},n\sigma}\hat c^\dagger_{j\sigma}\gsket$ (\ref{eqpsi}) with the matrix elements calculated by means of NRG, and $z=0.197$ as indicated by the dashed line in the left panel. Here $j$ is the site-index along the Wilson chain, and $j=-1$ corresponds to the $d$-orbital.  (We normalized $\psi_n$ to unity.) 
We compare this to the single-particle
eigenvectors of the quasiparticle Hamiltonian (\ref{eqqpschrod}), at the same quasiparticle weight. Results are shown for the four lowest quasiparticle states of the $L=29$ chain
that are indicated by arrows in the left panel. For comparison, we also show the single-particle wave functions of the non-interacting ($U=0$, $z=1$) chain, with the same bare hybridization $\Delta=0.09D$. We see near perfect coincidence between the NRG results and the single-quasiparticle wave functions at $z=0.197$. We also note that amplitudes
on low energy shells (large $j$) are nearly independent of $z$. The quasiparticle weight (or renormalized hybridization) only affects the very small amplitudes on energy shells
$j=0,2,\ldots$ up to the shell at the Kondo scale. This is a manifestation of the fact that the hybridization term in the Hamiltonian represents an irrelevant perturbation around the
infrared fixed point (albeit a leading one).

\begin{figure*}[htbp]
\begin{center}
\includegraphics[width=1.0\columnwidth]{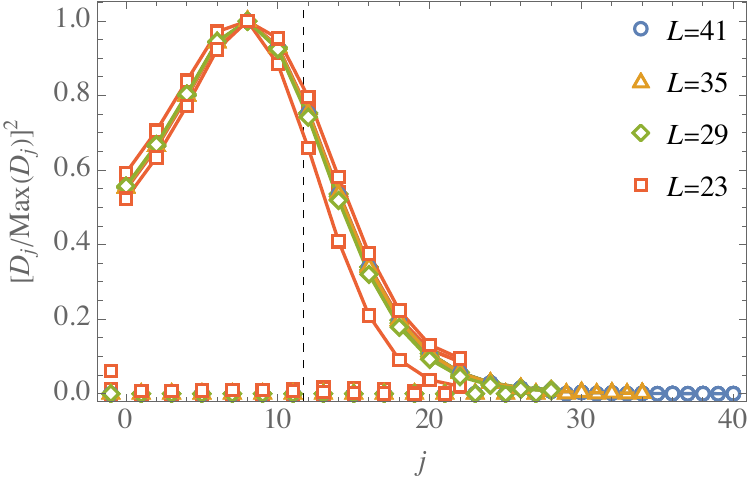}
\includegraphics[width=1.0\columnwidth]{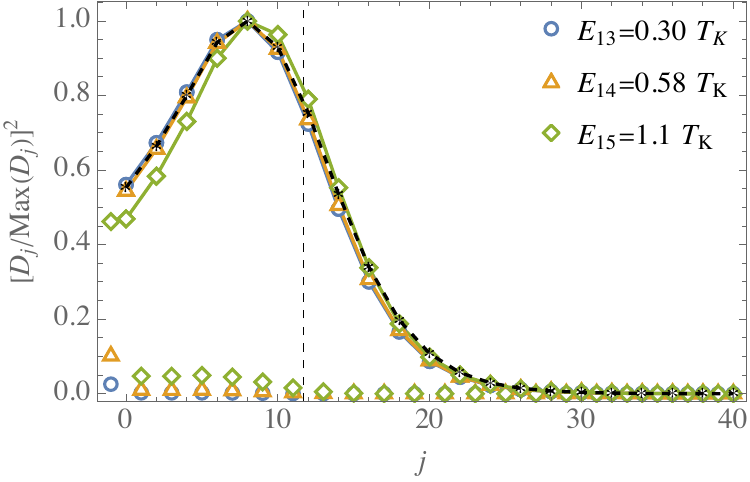}
\caption{{\bf Left panel:} NRG results for the dressing $D_j$, (\ref{eqdress}) normalized so that $\text{Max}(D_j)=1$, for the symmetric SIAM with  $\Delta=0.09 D$ and $U=0.6 D$.
Results were obtained with NRG on Wilson chains of lengths as indicated in the legend and $\Lambda=2$. At each length, the lowest four single particle states
are plotted. They lie on top of each other, except for $L=23$. The dashed vertical line indicates the position of Kondo temperature (\ref{eqtk}) relative to the energy shells $j$ of the 
Wilson chain.
{\bf Right panel:} Ansatz results for the dressing $D_j$, (\ref{eqdress}) normalized so that $\text{Max}(D_j)=1$, for the same symmetric SIAM as in the left panel.
Results were obtained on Wilson chains of length $L=41$. Different data sets correspond to different states calculated at the same $L$. Results are shown for the two
highest single quasiparticle states below $\tk$ and the lowest one above $\tk$. The black data represents the NRG result for the lowest single quasiparticle state at $L=41$.}
\label{fig3}
\end{center}
\end{figure*}

\subsection{Mode-resolved dressing.}

Next we present results for the mode-resolved dressing measure $D_{jn}$, cf. (\ref{eqdress}). We work on the Wilson chain, in the site basis $j$, i.e. we resolve dressing per energy shell. We have already investigated the total amount of dressing of a single-quasiparticle state, by calculating $p_n$. Here we
therefore normalize each $D_{jn}$ such that $\text{Max}\{D_{jn}|j=d,0,1,\ldots,L-1\}=1$. In the left panel of Figure~\ref{fig3}, we present NRG results for $|D_{jn}|^2$, for the
lowest four single-quasiparticle states $n=1,2,3,4$, for Wilson chains of different length. For chains long enough that the infrared cut-off scale is less than $\sim0.1\tk$,
we see that $D_{jn}$ has the same shape for $n=1,2,3,4$, and for different $L$. This is a non-trivial observation, since the associated $\bm y_n$, $n=1,2,3,4$ 
used to calculate $D_{jn}$
are clearly distinct. It means that low energy quasiparticles associated with different single-particle orbitals are dressed in the same form. 
We find that $D_{jn}$ is peaked at an energy shell above $\tk$. This reflects the fact that one has
to probe the system at ultraviolet scales to distinguish a quasiparticle from a bare electron. Regarding $L$-dependence, we 
see that $D_{jn}$ starts changing when the Wilson chain is so short that the tail of $D_{jn}$ extends to the lowest energy shell.
This behaviour confirms that the proposed measure really does give the correct mode-resolved picture: as the length $L$ of the Wison chain is decreased, the nature of quasiparticles starts changing when the low energy dressing cannot
fit on the Wilson chain any more. In the right panel of Figure~\ref{fig3}, we investigate the behaviour of single-quasiparticle excitations close to $\tk$ for a chain of length $L=41$, for which the infrared cut-off scale safely lies well below $\tk$. Results obtained with the ansatz are presented. The profile of $D_{jn}$ for low-energy excitaions obtained in the left panel via NRG is shown as a black dashed line. The two highest single-quasiparticle excitations below $\tk$ still fit this profile very well. For the lowest excitation above $\tk$ we see the profile starting to shift, and becoming significant on chain sites $j=1,3,\ldots$ where previously it was negligible. (For higher excited states, the local Fermi liquid picture rapidly fails.)    

\begin{figure*}[htbp]
\begin{center}
\includegraphics[width=.98\columnwidth]{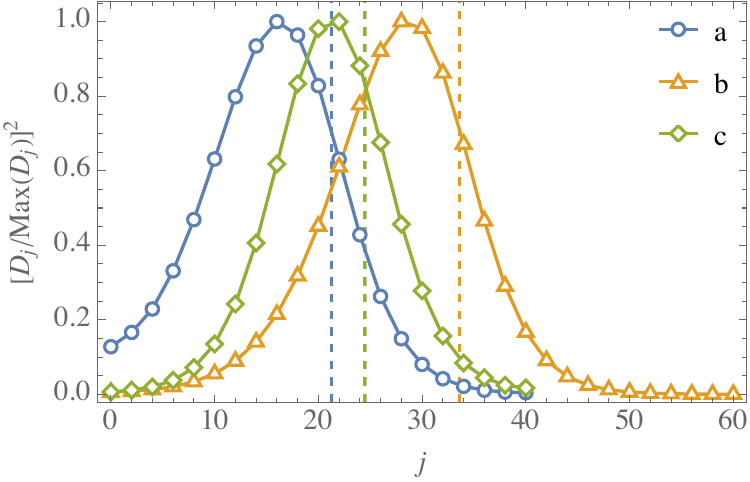}
\includegraphics[width=1.02\columnwidth]{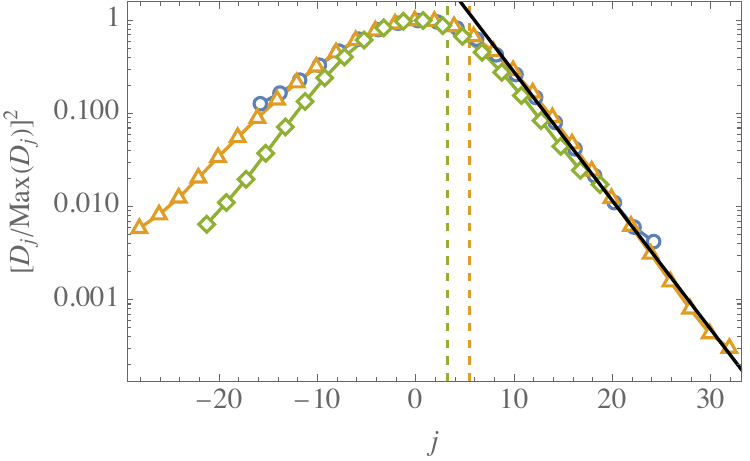}
\caption{{\bf Left panel}: Comparison of NRG results for the dressing $D_j$, (\ref{eqdress}) normalized so that $\text{Max}(D_j)=1$, for the symmetric SIAM for different
$\Delta$ and $U$. The three data sets are for (a) $\Delta=0.09D$, $U=1.4D$ (b) $\Delta=0.09D$, $U=2.4D$, and (c) $\Delta=0.0009D$, $U=0.006D$. The corresponding Kondo temperatures $\tk$ are respectively $4.7\times10^{-4}D$, $6.6\times10^{-6}D$, and $1.5\times10^{-4}D$.
The lowest state with a single quasiparticle on top of the ground state is plotted in each case.
Results were obtained with NRG on Wilson chains with $\Lambda=2$. The dashed vertical lines indicate the positions of Kondo temperatures 
(\ref{eqtk}) relative to the energy shells $j$ of the Wilson chain.
{\bf Right panel}: Same data as left panel, but plotted in log-scale on the vertical axis, and shifted horizontally so the maxima occur at zero. 
The black line is proportional to $e^{-j/\pi}$.}
\label{fig4}
\end{center}
\end{figure*}

We now return to the low-energy behaviour of (normalized) $D_{jn}$. In the left panel of Figure~\ref{fig3}, we show results for the lowest eigenstate with one quasiparticle on top of the ground state
($n=1$) for different combinations of $\Delta$ and $U$, all chosen such that $\tk\ll D$. We see that the general shape is rather similar for vastly different parameters. 
(Of course, if we did not normalize, the overall amplitude would depend strongly on $\tk$.) In the right panel, we replot the same data in log-scale on the vertical axis,
and shifted horizontally to line up the maxima of the different data sets at $j=0$. At small $j$ (i.e. in the ultraviolet) we see exponential behaviour that depends on $\Delta$, but
not on $U$. In the infrared (large $j$), we obtain universal behaviour $D_{jn}\sim e^{-j/2\pi}$. Translating from shell index to energy and then to momentum, 
this implies $D_{jn}\sim k^{1/\pi\text{ln}(\Lambda)}$, where $k$ measures momentum relative to the Fermi momentum. To extrapolate NRG results obtained with logarithmic discretization, to the original continuum system, one is supposed to take the $\Lambda\to1$ limit, but here the limit does not exist. 
It therefore seems that the low momentum- , or long-distance real space behaviour of the dressing profile $D_{kn}$, is beyond the reach of NRG. 

\subsection{Real-space quasiparticle structure}

\begin{figure*}[htbp]
\begin{center}
\includegraphics[width=.98\columnwidth]{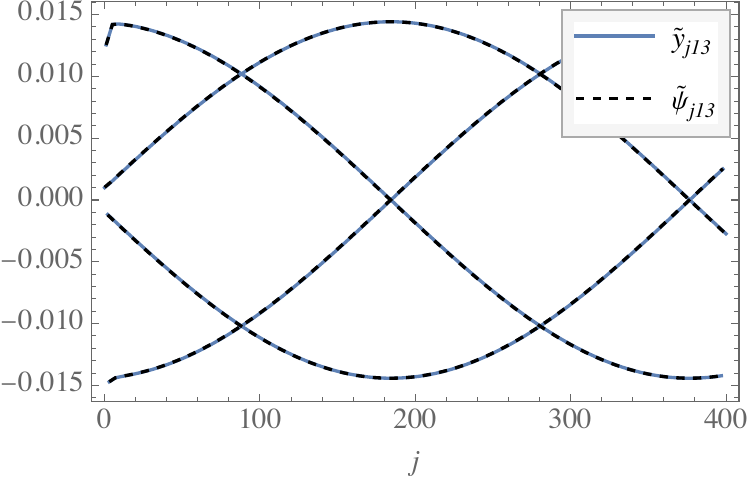}
\includegraphics[width=1.02\columnwidth]{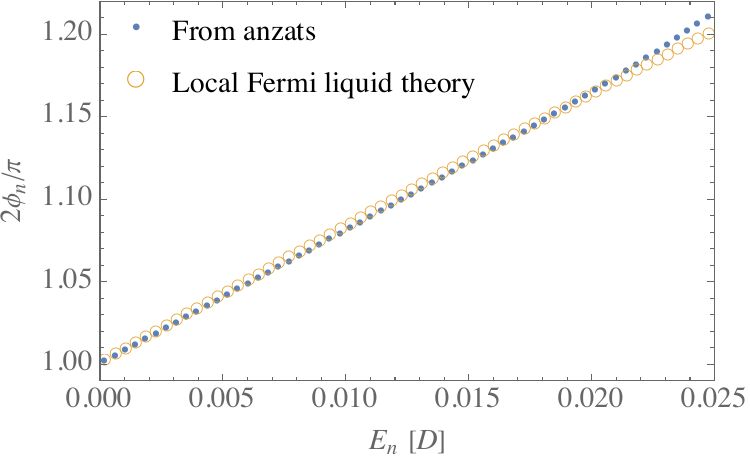}
\caption{{\bf Left panel:} A representative position space quasiparticle wave function (the 12'th excited single quasiparticle state) for the symmetric SIAM with  $\Delta=0.09 D$ and 
$U=0.2D$ calculated on a regular 1D lattice of 9601 sites, with the $d$-orbital coupled to the central site ($j=0$). The solid curves represents the result obtained with the Ansatz, while the dashed curves represent the corresponding eigenvector of the single-quasiparticle Hamiltonian (\ref{eqqpschrod}), with the quasiparticle weight estimated as indicated by the dashed line in the left panel of Figure (\ref{fig1}). Each curve was obtained by connecting amplitudes on every fourth site, starting at sites 0,1,2 or 3. {\bf Right panel:} The phase shift. Results extracted from the Ansatz, using (\ref{eqpsishifted}) are indicated by dots. The open disks were obtained from the single-quasiparticle Hamiltonian on the regular 1D lattice (\ref{eqqpschrod}), with the renormalized hybridization $\tilde V$ inferred from NRG calculations similar to those in Subsection~\ref{ssa}.}
\label{fig5}
\end{center}
\end{figure*}

To access the properties of single-quasiparticle excitations in real space, we therefore proceed to study the structure of single-quasiparticle states on a regular energy grid (\ref{eqlindiscr}) of $L=2\Omega+1$ levels with $\Omega=2400$, using natural orbital methods. 
We take $\Delta=0.09$ as above, and work at the relatively weak interaction strength $U=0.2D$. According to NRG, the quasiparticle weight 
$z=0.77$ and the Kondo temperature is $0.064D$. We have benchmarked the 
ansatz at this point in parameter space by calculating the Kondo resonance of the
local density of states of the $d$-orbital. We find that the ansatz is accurate for excitation energies up to $\sim 0.3 \tk$. (See Appendix~\ref{appa}.
For fixed size of the correlated sector, the ansatz on the large linear grid seems to be somewhat less accurate than on the logarithmic grid. On the linear grid and with
$U=0.6$ for instance, errors in the $10\%$ range are found for quantities such as the quasiparticle weight. This is why we chose to work with $U=0.2D$ here.) 

Previously, in the right panel of Figure~\ref{fig2}, we confirmed that matrix elements 
$\bra{\text{p},n\sigma} \hat d_{\sigma}^\dagger \gsket/\sqrt{z}$ and
$\bra{\text{p},n\sigma} \hat c_{k\sigma}^\dagger \gsket$ are eigenvectors of the single-quasiparticle
Hamiltonian (\ref{eqqpschrod}) for the Wilson chain. Now we perform the equivalent analysis for the large regular energy grid, to establish whether the equality
holds at single real-space lattice site resolution. 

In the left panel of Figure~\ref{fig5}, we plot $y_{kn}=\bra{\text{p},n\sigma} \hat c_{k\sigma}^\dagger \gsket$, 
Fourier transformed to real space, i.e.
\begin{equation}
\tilde y_{jn}=\frac{1}{\sqrt{2\Omega+1}}\sum_{k=1}^{2\Omega+1}
\cos\left[\frac{\pi j (2k-1)}{4\Omega+2}\right] y_{kn}.\label{eqfourier}
\end{equation} 
We can check this against the predictions of local Fermi liquid theory as follows. First we extract the renormalized $\tilde V$ by fitting the low energy NRG spectrum
to the Wilson chain version of the single-quasiparticle Hamiltonian (\ref{eqqpschrod}). Using this $\tilde V$, we compute the eigenvectors $\bm \psi_{n}$ 
of the single-quasiparticle Hamiltonian (\ref{eqqpschrod}) discretized on the regular energy lattice. We Fourier transform and suitably normalize the bulk amplitudes
to obtain $\tilde \psi_{jn}$. This is compared to $\tilde y_{jn}$. 
As a representative example, we look at the 12'th excited single-quasiparticle state (i.e. $n=13$), and see near-perfect agreement.

Except for the first few sites, $\tilde y_{jn}$ is a shifted sinusoidal wave
\begin{equation}
A_n\cos(\left|j\right|k_n-\phi_n). \label{eqpsishifted}
\end{equation}
We extract the phase shift $\phi_n$ by fitting (\ref{eqpsishifted}) to $\tilde y_{jn}$, 
and remembering that the boundary condition $\tilde y_{2\Omega+2,n}=0$ imposes the quantization
\begin{equation}
k_n=k_\text{F}+\frac{\pi n-\phi_n}{2\Omega+1}, \label{eqkquantization}
\end{equation}
of the wave number.
We can compare this to the predictions of local Fermi liquid theory by similarly extracting the phase shift of  $\tilde \psi_{jn}$.
In the right panel of Fig.~\ref{fig5}, we see that the
two estimates agree very well up to $E_n=0.02D\sim0.3\tk$. 
We conclude that indeed the matrix elements $y_{dn}/\sqrt{z}$ and $\tilde y_{jn}$ represent a real-space eigenvector of the single-quasiparticle
Hamiltonian (\ref{eqqpschrod}). 

If a non-interacting quasiparticle picture holds, we should equivalently be able to extract the phase shift from the quantization of the
wave number and the dispersion relation as follows 
\begin{equation}
E_n=\frac{2\Omega+1}{\pi\Omega}(k_n-k_\text{F})\implies \phi_n=\pi\left(n-\frac{\Omega E_n}{D}\right). \label{eqpsfrome}
\end{equation}
We have checked that applying this method to results obtained with the ansatz gives the same result as the dots in the right panel of Figure~\ref{fig5}.
 
\begin{figure}[htbp]
\begin{center}
\includegraphics[width=\columnwidth]{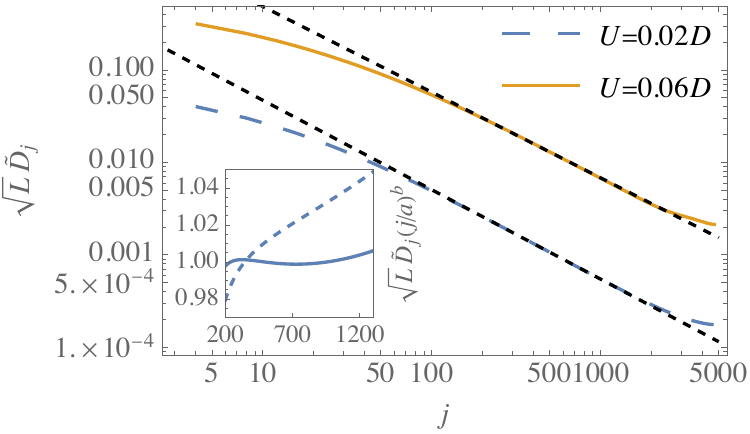}
\caption{{\bf Main panel:} The dressing measure $D_j$ Fourier transformed to real space as in (\ref{eqfourier}), for the symmetric SIAM calculated on a regular 1D lattice of 9601 sites, with the $d$-orbital coupled to the central site ($j=0$) at $\Delta=0.09 D$ and two values of $U$. 
The curves are envelopes to the actual data, and were obtained by plotting the result on each fourth site, $j=4l$, $l=0,1,\ldots$.
In each case the lowest single-quasiparticle state as obtained by the Ansatz, was used to calculate 
$D_j$. The dashed lines represent a fit to a power law $(a/j)^b$. For $U=0.2$ we found a best fit $b=0.97$ and for $U=0.6$ we found $b=0.93$. {\bf Inset}: The solid curve
represents $\sqrt{L} \tilde D_j \times (j/a)^b$ for $U=0.2D$, with $a=0.43$ and $b=0.97$ determined in the main panel by fitting $(a/j)^j$ to $\sqrt{L} \tilde D_j$.
The dashed curve
represents $\sqrt{L} \tilde D_j \times (j/a)$ for $U=0.2D$, with $a=0.53$ determined by fitting $a/j$ to $\sqrt{L} \tilde D_j$. From this we deduce that the data is sufficient to 
reveal that the powerlaw is $\tilde D_j\sim j^{-0.97}$ rather than $\tilde D_j\sim j^{-1}$ at $U=0.2D$.}
\label{fig6}
\end{center}
\end{figure}

Finally, we study the dressing profile in real space. That is, we compute $D_{kn}$ for the system discretized on a regular energy grid, and Fourier transform as $(\ref{eqfourier})$
to obtain the site-resolved dressing $\tilde D_{jn}$. The results presented above, particularly in the left panel of Figure~\ref{fig4}, suggests that $\tilde D_{jn}$ is a good measure of
how different from a bare electron the quasiparticle excitation looks on real-space lattice site $j$.
In Figure~\ref{fig6}, we show results at $U=0.2D$, which are quantitatively accurate, and results at $U=0.6D$, that may contain small but non-negligible errors. 
We compare the lowest eigenstate with one quasiparticle on top
of the ground state in each case, for which the ansatz should give more accurate results than for higher excited states. We therefore think that the error in the $U=0.6D$ results are 
smaller than the $10\%$ errors observed for quantities such as the quasiparticle weight, that tests the ansatz at higher excited states. As expected we see that the dressing decays as we move away from the impurity, and that increasing $U$, which decreases $\tk$, leads to more significant dressing. At large distances, the dressing profile seems to satisfy a power law $\sim j^{-b}$ with $b$ discernibly different from $1$. For $U=0.2D$, we find $b=0.97$ while for $U=0.6D$ we find $b=0.93$. The decrease of the exponent at larger $U$ is consistent with the expectation that the particle-hole pairs that dress the bare electron are less confined, the lower the Kondo temperature, but it has to be kept in mind that the $U=0.6D$ results are less accurate than the $U=0.2D$ results. We are therefore less confident about the exact value of the exponent at $U=0.6D$ than at  $U=0.2D$.
Nonetheless, evidence of a nontrivial power law is clear, and could not have been obtained with any method other than the ansatz, as far as we can see.

\section{Summary and conclusion}
\label{sec5}
There are two distinct single-particle-like structures associated with dynamic fermionic quantum impurity problems. There are the quasiparticles
that provide a local Fermi liquid description of low energy excitations in terms of independent {\it effective} degrees of freedom. Then there is also the natural orbital basis, associated with the {\it bare} electronic degrees of freedom, which is known to provide an economical description of ground state correlations. The work presented here was inspired by the
question: ``Can a synergy of these two pictures shed light on how bare electrons are dressed to form the quasiparticle excitations of a local Fermi liquid?''. 
We succeeded in establishing the following. The quasiparticle weight $z$, and the effective quasiparticle Hamiltonian 
\begin{equation}
\hat H_\text{qp}=\sqrt{z}V\sum_\sigma\left(\hat d^\dagger_\sigma \hat \psi_{0\sigma}+\hat \psi_{0\sigma}^\dagger \hat d_\sigma\right)
+\sum_{k=1}^L\sum_\sigma \varepsilon_k \hat c_{k\sigma}^\dagger \hat c_{k\sigma},\label{eqqpham}
\end{equation} 
of the particle-hole symmetric Single Impurity Anderson Model have meaning not only in the infinite system, where they determine the scattering matrix of electrons at low energy,
but also in the large finite system. If $\ket{\text{p},n\sigma}$ is a low energy single-quasiparticle excited state of the SIAM, then to a good approximation
$\psi_{dn} =\bra{\text{p},n\sigma} \hat d^\dagger_{\sigma}\gsket/\sqrt{z}$ and $\psi_{kn} =\bra{\text{p},n\sigma} \hat c_{k\sigma}^\dagger\gsket$ 
solve the single-particle Schr\"odinger equation (\ref{eqqpschrod}) associated with (\ref{eqqpham}).
Furthermore, if we define $y_{dn}=\sqrt{z/p_n}\psi_{dn}$ and $y_{kn}=\psi_{kn}/\sqrt{p_n}$ with $p_n$ such that $(y_{dn},y_{1n},\ldots,y_{Ln})$ is normalized to unity,
then $\hat a_{n\sigma}^\dagger= y_{dn}\hat d_{\sigma}^\dagger+\sum_{k=1}^L y_{kn}\hat c_{k\sigma}^\dagger$ represents the best bare electron approximation to 
the quasiparticle associated with $\ket{\text{p},n\sigma}$. We call $p_n$ the bare electron occupation probability because it is the occupation probability of the 
$\hat a_{n\sigma}$-orbital when the system is in state $\ket{\text{p},n\sigma}$. In a system where the single-quasiparticle level spacing is $\delta E$ in the vicinity of excitation
$\ket{\text{p},n\sigma}$, at low energies, $p_n$ is nearly independent of energy and related 
to the quasiparticle weight $z$ by $z=1/(1+p_n \pi\Delta/\delta E)$. We define the dressing of the bare electron as
$\ket{\text{dressing}}=\ket{\text{p},n\sigma}-\sqrt{p_n}\hat a_{n\sigma}^\dagger\gsket$, and obtain a mode-resolved measure $D_{kn}$ of dressing by taking the overlap with 
$\hat c_{k\sigma}^\dagger\gsket$. This measure can be computed from knowledge of $p_n$, $(y_{dn},y_{1n},\ldots,y_{Ln})$, and the ground state covariance matrix $C(\gsket)$
whose eigenvectors define the natural orbital basis.
Where possible, we used NRG to study these quasiparticle-structure-related quantities. However, NRG is inadequate to study more than
the first few single-quasiparticle excitations, or to study structure in real space.  We overcame these limitations
by constructing an ansatz for single-quasiparticle excitations in terms of the natural orbital basis. 
The important features of this ansatz are (1) that it provides an explicit expression in terms of bare electronic degrees of freedom, for a 
quasiparticle excitation and (2) that it can be implemented directly on a microscopic model involving an impurity coupled to electrons on a lattice of several thousand sites, rather than on a continuum model rediscretized on the logarithmic energy grid used in NRG. The ansatz is accurate at affordable computational cost
in a regime of well-developed correlations. We further showed that the difference $\delta\rho$ between the density of single-particle states of $\hat H_\text{qp}$ and the actual
many-body system is an increasing function of energy at energies approaching or equal to the Kondo temperature. This is a ``beyond ideal Fermi-liquid theory'' effect.
We finally showed that the position resolved dressing $\tilde D_{jn}$ decays with a nontrivial powerlaw at large distances. 

We envisage that the work we presented here could provide a foundation for the explicit calculation of the single particle scattering matrix for a quantum impurity imbedded in
a non-trivial geometry such as a disordered host or a mesoscopic electronic device. This is a challenging task because the quasiparticle weight $z$ which appears in 
the effective Hamiltonian, is determined by
the many-body state of the system, which cannot be calculated with methods such as NRG, with its crudely resolved discrete representation of the host.   
In the present work, we considered single quasiparticle excitations. The methods we developed
may be a good starting point to investigate the weak interactions between two or more quasiparticles. Finally, we also suspect that simple modifications to our ansatz could extend its regime of validity. For instance, in the present work, we determine the natural orbital basis at the start of the calculation, using the ground state as a reference. In our own ongoing
work, we are investigating whether it is profitable to adjust the natural orbital basis itself when dealing with excitations.      

\begin{acknowledgements}
Insigthful comments and suggestions from Serge Florens are gratefully acknowledged.
\end{acknowledgements}

\appendix
\section{Benchmarking}
\label{appa}

We benchmark our proposed methods against known
results. To gauge generality, we test the natural orbital methods on the SIAM and on another model. The choice of this model
is dictated by the demand that it should be sufficiently different from the SIAM to give independent confirmation of the accuracy of the method, 
yet its structure must be similar enough that computer codes developed
for the SIAM can easily be adapted. 

\subsection{The Interacting Resonant Level Model}

The Interacting Resonant Level Model (IRLM) 
is ideal for this purpose. It describes a band of non-interacting, spinless fermions hybridizing with a localized
orbital. Additionally there is a local interaction between a particle in the local orbital and one in the adjacent 
lattice site of the band. We study the particle-hole symmetric version of the Hamiltonian, which reads 
\begin{align}
\hat H=&U\left(\hat n_d-\frac{1}{2}\right)\left(\hat n_0-\frac{1}{2}\right)
+V\left(d^\dagger \psi_0+\psi_0^\dagger d\right)\nonumber\\
&+\sum_{k=1}^L\varepsilon_k \hat c_k^\dagger \hat c_k.\label{eqhamirlm}
\end{align}
Here $\hat d^\dagger$ creates a particle in the localized orbital and $\hat \psi_0^\dagger$ creates a particle on the lattice site adjacent to the 
impurity. Furthermore 
\begin{equation}
\hat n_d=\hat d^\dagger \hat d,~~~\hat \psi_0=\frac{1}{\sqrt{L}}\sum_{k=1}^L\hat c_k,~~~\hat n_0=\psi_0^\dagger \psi_0.\label{eqirlmops}
\end{equation}
We take the Fermi energy to be at zero, in the middle of the band. 
As in the SIAM, hybridization is quantified by the spectral density $\Delta=\pi V^2/2D$ at the Fermi level, of the operator $V\hat\psi_0$ in the infinite system
uncoupled from the impurity.
In the $\Delta\ll D$ regime, and for $U<0$, the IRLM hosts the same physics as the Kondo model, in which there is a local exchange interaction between the 
magnetic moment of a spin-1/2 impurity and nearby conduction band electrons. (There is a mapping between the spin degrees of freedom in the Kondo model
and the IRLM, the charge degrees of the Kondo model being spectators that do not couple to the impurity \cite{Guinea,Kotliar,CostiZarand}.) The quantum phase transition seen in the
Kondo model when the exchange interaction switches from ferro- to antiferromagnetic, translates to a critical point in the IRLM at $U\sim -D$. 
Increasing $U$ beyond the critical point corresponds to increasing the Kondo temperature in the antiferromagnetic regime.

\subsection{Ansatz 2} 

Apart from the study of Fermi liquid quasiparticles, we are also interested in establishing if there are any fundamental limitations to applying
natural orbital methods to excited state problems. 
Even though the ansatz formulated in Section~\ref{sec3}
will prove sufficiently accurate for our purposes, we introduce an improved, but more computationally expensive Ansatz 2 for the purpose of comapring
to the ansatz of Section~\ref{sec3}, which here we refer to as Ansatz 1.

Ansatz 2 is obtained by relaxing the assumption that when a particle ends up in the unoccupied sector, the particles in the correlated sector are undisturbed.
Instead, a particle in the unoccupied sector may be associated to an arbitrary configuration of $m$ particles in the correlated sector.
Thus we consider $N+1$ particle states of the form 
\begin{eqnarray}
\ket{\text{p},k}_2&=&\sum_{\alpha\in\mathcal U}\sum_{\sset{\alpha}{m}}u_{\alpha,\sset{\alpha}{m},k}\,\hat q_\alpha^\dagger\ket{\sset{\alpha}{m}}\nonumber\\
&&+\sum_\sset{\alpha}{m+1}v_{\sset{\alpha}{m+1},k}\ket{\sset{\alpha}{m+1}}.\label{eqansatz2p}
\end{eqnarray}
The optimal coefficients $u_{\alpha,\sset{\alpha}{m},k}$ and $v_{\sset{\alpha}{m+1},k}$ are found in the same manner as before. They are thus seen to be
eigenvectors of the effective Hamiltonian
\begin{equation}
\tilde H_\text{p}=\left(\begin{array}{cc}\tilde H_{\mathcal U\mathcal U} &\tilde H_{\mathcal U\mathcal C}\\
\tilde H_{\mathcal U\mathcal C}^\dagger&H_{\mathcal C\mathcal C}
\end{array}
\right),
\label{eqhammat2}
\end{equation}
where the blocks $\tilde H_{\mathcal U\mathcal U}$ and $\tilde H_{\mathcal U\mathcal C}$ are now given by
\begin{eqnarray}
\left[\tilde H_{\mathcal U \mathcal U}\right]_{\left(\alpha,\sset{\alpha}{m}\right),\left(\beta,\sset{\beta}{m}\right)}&=&\bra{\sset{\alpha}{m}} \hat q_\alpha \hat H \hat q^\dagger_\beta
\ket{\sset{\beta}{m}},\nonumber\\
\left[\tilde H_{\mathcal U \mathcal C}\right]_{\left(\alpha,\sset{\alpha}{m}\right),\sset{\beta}{m+1}}&=&\bra{\sset{\alpha}{m}} \hat q_\alpha \hat H 
\ket{\sset{\beta}{m+1}},\nonumber\\
\label{eqhamblocks2}
\end{eqnarray}
while the block $H_{\mathcal C\mathcal C}$ is the same as before (\ref{eqhamblocks1}). The block $\tilde H_{\mathcal U\mathcal U}$ has dimension 
$N_{\mathcal U}\times\left(\begin{array}{c} N_{\mathcal C}\\ m\end{array}\right)$, which is considerably larger than the corresponding block (\ref{eqhamblocks1}) in the 
effective Hamiltonian associated with Ansatz 1.

\subsection{Local density of states of the $d$-orbital.}

We will explore the accuracy of Ansatz 1 and Ansatz 2 by considering the the local density of states (LDOS) of the $d$-orbital, a quantity that is sensitive to 
single-particle excitations on top of the ground state. The two ans\"atze that we study were designed to be accurate in the Fermi liquid regime, i.e. below the
Kondo scale. We intentionally choose the LDOS as a benchmark, not only to demonstrate the accuracy of the ans\"atze in the expected regime, {\it but also
to explore how they break down}. We will see that the LDOS is accurately reproduced at frequencies up to $\sim \tk$, and start breaking down at higher
frequencies, where the local Fermi liquid picture does not apply. We stress that the ans\"atze studied here, are not being presented to compete with existing methods
to compute the LDOS of the $d$-orbital. Instead, they come into their own right when quasiparticles living in the bulk, rather than impurity properties, are considered.      

For a finite system, in which the band has $L$ orbitals, the LDOS of the $d$-orbital is given by
\begin{eqnarray}
A_L(\omega)&=&\frac{1}{\pi}\text{Re}\,\int_0^\infty dt\, e^{i\omega t}\gsbra\left\{\hat d_\sigma^\dagger(0),\hat d_\sigma(t)\right\}
\gsket\nonumber\\
&=&\sum_{n}\left|\gsbra\hat d^\dagger_\sigma\ket{\text{h},n\sigma}\right|^2\delta(E_{\text{h},n}-E_\text{GS}+\omega)\nonumber\\
&&+\sum_{n}\left|\gsbra\hat d_\sigma \ket{\text{p},n\sigma}\right|^2\delta(E_{\text{p},n}-E_\text{GS}-\omega).\nonumber\\
\label{eqdos}
\end{eqnarray}
For the SIAM, we assume zero magnetic field, so that a single spin species may be considered. For the IRLM, the spin index $\sigma$ is dropped.
The summations in respectively the second and third lines are over all excited states with one more or fewer particles than the number of particles in the ground state.
For results obtained in the logarithmically discretized model, we use a trick called  $z$-averaging to mitigate discretization errors at finite frequency \cite{Oliveira}.
It involves replacing $\Lambda^{-k}$ in (\ref{eqshells}) with $\Lambda^{-k+z}$ for $k=1,2,\ldots$, and averaging LDOS results over $n_z$ separate calculations,
each performed with a different $z=(\zeta-1/2)/n_z$, $\zeta=1,2,\ldots,n_z$. Note that with this discretization, the hopping amplitudes $t_j$ along the Wilson
chain (\ref{eqchain}) have to be calculated recursively \cite{CostiRMP}, using arbitrary precision arithmetic.  
 
In order to arrive at a result independent of the discretization scheme, and that is near the LDOS in the thermodynamic limit,
the delta-peaks of the finite system LDOS must be broadened. We apply the Gaussian broadening prescription that
is standard within NRG 
\begin{align}
&A(\omega)=\int_{-\infty}^\infty d\varepsilon A_L(\varepsilon)\Gamma(\omega,\varepsilon),\nonumber\\
&\Gamma(\omega,\epsilon)=\frac{1}{\sqrt{2\pi\eta(\varepsilon)}}e^{-(\omega-\epsilon)^2/2\eta(\varepsilon)^2}.
\label{eqbroadening}
\end{align}
For results obtained in the logarithmically discretized model, we use a
width that decreases linearly as the Fermi energy is approached, 
\begin{equation}
\eta_{\text{log}}(\varepsilon)=\frac{2|\varepsilon|}{n_z}.\label{eqetalog}
\end{equation}
For results obtained in the system discretized on a regular energy grid, we use a constant width equal to half the single-particle level spacing
\begin{equation}
\eta_{\text{lin}}(\varepsilon)=\frac{D}{L}.\label{eqetalin}
\end{equation}
If an STM tip is held near the $d$-orbital and a potential $\omega/e$ is maintained
between tip and sample, the tunnelling current between the tip and the $d$-orbital is proportional to $A(\omega)$.
We note that the LDOS of the IRLM is still a topic of active study \cite{Camacho}.

Here and further below, all NRG results for the LDOS of the $d$-orbital, were obtained with the density matrix NRG (DM-NRG) algorithm \cite{Hofstetter} 
which retains full information of the ground state when the LDOS
is evaluated at an arbitrary frequency. In order to use the maximum available information about excited states, we employ the Full Hilbert space method \cite{Peters}.
Thus the sum rule $\int_{-\infty}^\infty d\omega\,A_L(\omega)=1$ is satisfied identically. 
We imposed particle number conservation during the iterative diagonalization, and retained up to 128 states per particle-number sector, which corresponds to several thousand
kept states in total per iteration.  

\subsection{Results for the IRLM}

\begin{figure}[htbp]
\centering
\includegraphics[width=1.0\columnwidth]{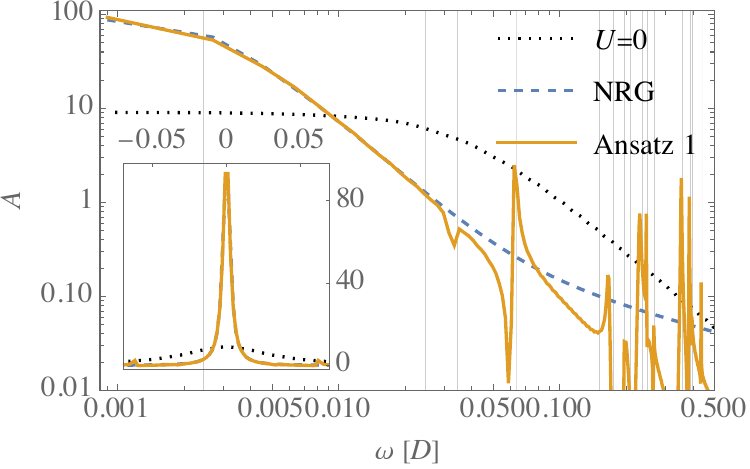}
\includegraphics[width=1.0\columnwidth]{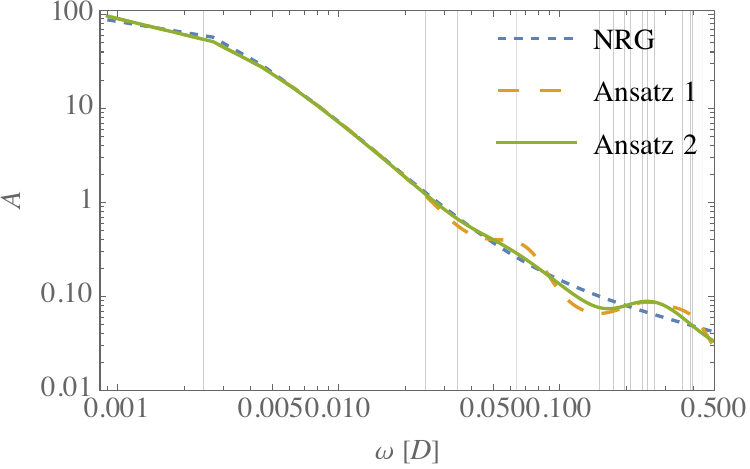}
\vspace{-0.5cm}
\caption{{\bf Top panel:} Local Density of States (LDOS) at the $d$-orbital, for the Interacting Resonant Level Model (IRLM), at interaction strength $U=-0.5D$ and hybridization $\Delta=0.035D$.
The main panel shows results in log-log scale. The inset shows the same data in linear scale. The solid curve represents the result obtained with Ansatz 1. The band was
discretized on a regular energy grid with 4801 levels. The dashed curve represents an NRG calculation. The dotted curve represents the non-interacting ($U=0$) result for comparison. Vertical lines indicate the eigenvalues of $H_{\mathcal C \mathcal C}$ (\ref{eqhamblocks1}).
{\bf Bottom panel:} LDOS of the IRLM at the same parameters $U=-0.5D$ and hybridization $\Delta=0.035D$ as in the top panel. The NRG data is the same as in the top panel. 
Results for Ansatz 2 (solid curve) were obtained using the same logarithmic discretization and $z$-averaging as the NRG results. For comparison, results for Ansatz 1 were also recalculated on the logarithmic grid, with  $z$-averaging.}  
\label{figa1}
\end{figure}

First we present results for the IRLM.
We calculate $A(\omega)$ at $\Delta=0.035D$ and $U=-0.5D$. We compare NRG results to results obtained with Ansatz 1 on a regular energy grid of 4801 
orbitals. Natural orbitals were calculated using the RGNO algorithm, with a correlated sector containing four particles in $N_{\mathcal C} = 8$ orbitals. 
NRG results were obtained on a Wilson chain of length $52$ sites, including the $d$-orbital, with $\Lambda=1.5$, leading to an infrared cutoff energy $\sim 4\times 10^{-5} D$. NRG results were $z$-averaged over 8 $z$-values.
Results are shown in the top panel of Figure~\ref{figa1}, in logarithmic scale, and in linear scale (inset). 
The peak around $\omega=0$ is the well-known Kondo resonance. 
The non-interacting case, $U=0$, corresponds to the Toulouse point of the anisotropic Kondo model, and therefore still 
shows a Kondo resonance (of half-width $\Delta=0.035D$). We see that at $U=-0.5 D$, the resonance is about 10 times narrower. This significant downward 
renormalization of the hybridization proves that we are in the strongly correlated regime. We see that Ansatz 1 implemented on a regular energy grid reproduces the 
LDOS well, from the Fermi energy up to the point where it has decayed to about 1\% of its value at the Fermi energy. It therefore captures low-energy excitations well,
even beyond the Kondo scale (half-width of the Kondo resonance). The eventual breakdown occurs as follows.
The effective Hamiltonian (\ref{eqhammat1}) describes the hybridization of a (near) continuum of bare single-particle excitations in the unoccupied sector with a discrete spectrum of few-body correlated states in the correlated sector. This discrete spectrum comprises the eigenvalues of the lower-right block $H_{\mathcal C\mathcal C}$ of $H_\text{p}$ in (\ref{eqhammat1}) (together with the eigenvalues of the particle-hole conjugate, representing one-hole excitations). In Figure~\ref{figa1}, we plot vertical lines representing
this discrete spectrum. We see that Ansatz 1 hybridizes the two few-body states that have energies closest above and below the Fermi energy, 
with the continuum of bare particle and bare hole excitations, to produce a smooth
spectral density in the vicinity of the Kondo resonance. The remaining few-body correlated states have energies that lie in the tail of the Kondo resonance. Ansatz 1 does not
produce the correct hybridization of these states with the bare particle and hole continua, as seen from the abrupt features of the resulting spectral density in the vicinity
the vertical lines in the tails of the Kondo resonance. While this analysis indicates that Ansatz 1 is sufficiently accurate to allow us to investigate the structure of local Fermi-liquid quasiparticles, we would nonetheless like to see if we cannot improve accuracy at higher frequencies. 

As we explained above, Ansatz 2 is too computationally expensive to deploy on a large regular energy grid.  
In the bottom panel of Figure~\ref{figa1}, we compare Ansatz 1 and Ansatz 2 to NRG results, where now all calculations are performed for the IRLM discretized on 
the logarithmic grid, and then $z$-averaged. We again use the RGNO algorithm, involving a correlated sector of $N_{\mathcal C} = 8$
orbitals, containing 4 particles. 
The interaction strength $U$ and hybridization $\Delta$ are the
same as in the top panel. As might be expected, $z$-averaging, together with the increased broadening (\ref{eqetalog}) at intermediate energies, compared to (\ref{eqetalin}), smoothes the spurious abrupt features in the LDOS obtained by means of Anzatz 1. Nonetheless, in the bottom panel of Figure~\ref{fig1}, Ansatz 1 still shows deviations from the NRG result that stand in one to one correspondence with those seen in the top panel, for 
$|\omega|>0.025 D$. Ansatz 2 significantly mitigates the lowest frequency deviations and remains accurate up to $|\omega|\sim 0.1D$, describing the LDOS of the
IRLM well over almost three decades. The benchmarking exercise thus proves that natural orbital methods are in principle suitable to study low energy excitations of fermionic 
quantum impurity models. We therefore proceed to our primary system of interest, the Single Impurity Anderson Model (SIAM).

\subsection{Results for the SIAM}

As in Section~\ref{sec5}, we study the 
particle-hole symmetric version of the model (\ref{eqhamsiam}) where $\varepsilon_d=-U/2$. 
The LDOS of the $d$-orbital shows a resonance with a width $\tilde \Delta = z \Delta$, comparable to $\tk\ll U$, associated with spin fluctuations. 
Without the Coulomb repulsion $U\hat n_{d\uparrow} \hat n_{d\downarrow}$, this Kondo resonance is absent, and the LDOS has
two peaks at $\omega=\pm \varepsilon_d$ instead. In the presence of Coulomb repulsion, these turn into shoulders at the base
of the Kondo resonance, or even (broad) side peaks, 
at a scale $\varepsilon_d=-U/2$ and $U+\varepsilon_d=U/2$, 
associated with charge fluctuations. We take $\Delta=0.09 D$ and $U=0.6 D$, for which we compute  $\tk=0.013D$.
We perform all calculations on a logarithmic grid, and employ $z$-averaging over $n_z=8$ values of $z$. 
We employed a Wilson chain of length $L=51$ sites, with $\Lambda=2$. Natural orbitals were obtained with the
RGNO algorithm. A correlated sector with $N_{\mathcal C}=12$ orbitals (6 spin up and 6 spin down), containing $3$ spin up and $3$ spin down electrons
was used. The RGNO algorithm could determine the ground state energy to an error of $\sim 5\%$ of the Kondo temperature, meaning that we may expect 
Kondo physics to be described accurately. We note that the accuracy of the ground state calculation is easily improved by increasing the size of the correlated
sector. However, the enlarged correlated sector significantly slows down the subsequent calculation of excited states. We have therefore settled for the 
minimum accuracy that is sufficient for the excited state calculation. 

\begin{figure}[htbp]
\centering
\includegraphics[width=1.0\columnwidth]{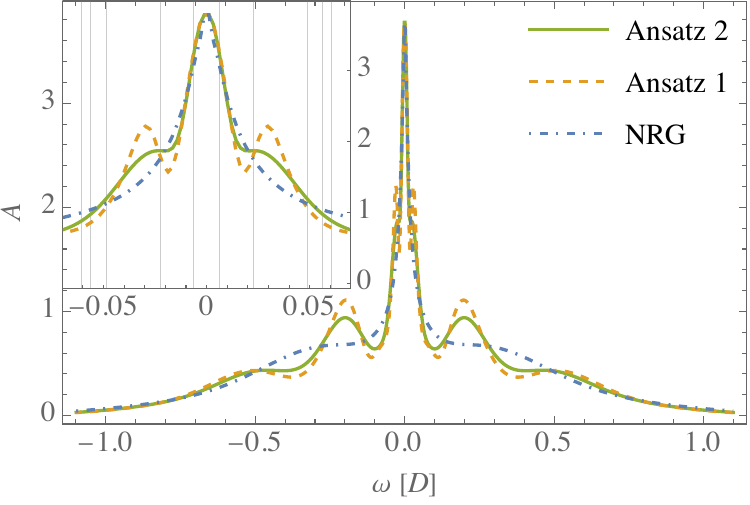}
\vspace{-0.5cm}
\caption{The local density of states (LDOS) at the $d$-orbital, for the symmetric Single Impurity Anderson Model (SIAM) with $\Delta=0.09 D$ and $U=0.6 D$. 
The main panel shows the result over a frequency window that spans the whole band. The inset zooms in on the Kondo resonance. The vertical lines in the inset represent the eigenvalues of  $H_{\mathcal C \mathcal C}$ (\ref{eqhamblocks1}). Results for Ansatz 1 and Ansatz 2 were obtained employing the same logarithmic discretization 
and $z$-averaging as in NRG.}
\label{figa2}
\end{figure}
In Figure~\ref{figa2} we compare NRG results for the LDOS of the $d$-orbital, to results obtained using Ansatz 1 and Ansatz 2. The NRG results show a clear Kondo
resonance of half-width $\sim 0.02D$ and broad shoulders associated with the scale $U/2=0.3D$. Both Ansatz 1 and Ansatz 2 reproduce the general shape well and
are quantitatively accurate at low frequencies. The inset to Figure~\ref{fig2} shows a zoom of the central peak of the LDOS, with vertical lines indicating the eigenenergies
of $H_{\mathcal C, \mathcal C}$  (\ref{eqhamblocks1}). As in the case of the IRLM, Anzatz 1 and Anzatz 2 both handle the hybridization of the two few-body correlated eigenstates of $H_{\mathcal C, \mathcal C}$ that are closest in energy to the Fermi level well (the two vertical lines at $\pm 0.006D$). Again, deviations from NRG results are associated with higher excited few-body correlated states, for instance those represented by the vertical lines at $\pm 0.02 D$. Ansatz 2 mitigates these deviations compared to Ansatz 1, but errors at this frequency scale are still on the order of 5\% for Ansatz 2. 

\begin{figure}[htbp]
\begin{center}
\includegraphics[width=\columnwidth]{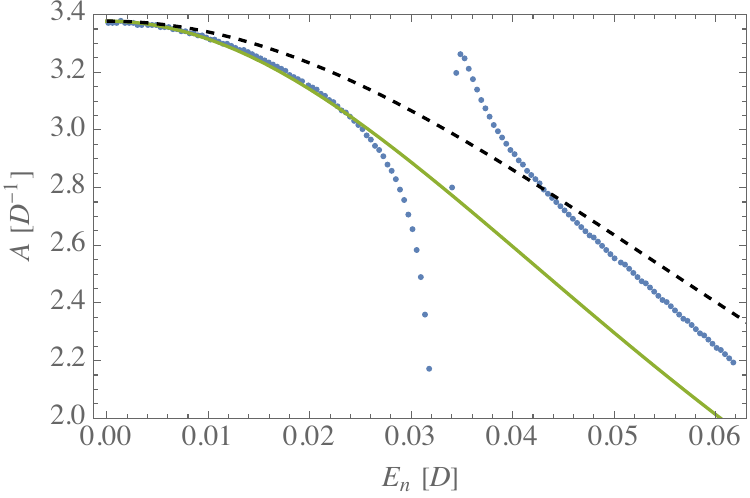}
\caption{The LDOS of the $d$-orbital (\ref{eqdos}) as calculated with the Ansatz (dots), for the symmetric SIAM with  $\Delta=0.09 D$ and $U=0.2 D$, calculated on a regular energy
grid with $L=4801$ orbitals. 
The solid line represents the Lorentzian LDOS  $z^2\Delta/\pi(\omega^2+z^2\Delta)$, with $z$ estimated  
from the low-energy spectrum obtained with NRG. The Dashed line represents the unrenormalized $z=1$ Lorentizain LDOS at the same $\Delta$. }
\label{figa4}
\end{center}
\end{figure}

At this point we conclude our study of Ansatz 2. Benchmarking shows that Ansatz 1 is accurate in the Fermi liquid regime below $\tk$.
It remains to check Ansatz 1 for the SIAM discretized on the large regular energy grid necessary for obtaining real space resolution at the scale of the Fermi wavelength.
Here we find that results are less accurate than on the Wilson chain. To obtain a quantitatively accurate LDOS on a regular energy grid with $4801$ levels
at $\Delta=0.09$, we had to reduce the interaction strength to $U=0.2D$. This gives a quasiparticle weight $z=0.77$, which still represents a finite renormalization.
In Figure~\ref{figa4} we show the low-energy LDOS computed with the anzatz, compared to what it should be for $z=0.77$. We see good agreement up to $\omega=0.02D$,
and a breakdown at $\omega=0.03D$. Here $\tk=0.064D$. From various trial calculations we performed with the ansatz, we believe that the energy range in which the ansatz is
accurate would be extended, if we could use a larger correlated sector, but that would make the calculation too expensive computationally, to perform in a reasonable time.

\bibliography{localfermiliquid}
\end{document}